\documentclass[twocolumn,apj,numberedappendix,twocolappendix]{openjournal}
\usepackage{graphicx}	% Including figure files
\usepackage{amsmath}	% Advanced maths commands   
\usepackage[dvipsnames]{xcolor}  
\usepackage{amssymb}	% Extra maths symbols
\usepackage[T1]{fontenc}
\usepackage{newtxtext,newtxmath}    % PUT THIS ONE BACK IN IN OVERLEAF
\usepackage{float}
\usepackage{ulem}
\usepackage{soul}
\usepackage{makecell}
\usepackage{multirow}
\usepackage{fontawesome}

\setcitestyle{aysep={}}

\usepackage{lineno}
\usepackage{orcidlink}

\usepackage{caption}
\usepackage{natbib}
\usepackage{macros_vdb} % my personal macros

\usepackage{hyperref}
\hypersetup{colorlinks=true, linkcolor=blue, citecolor=blue, filecolor=black, urlcolor=blue}

\begin{document}

\title{The Impact of Baryonic Effects on the Dynamical Masses inferred using Satellite Kinematics}
      
\shorttitle{The Impact of Baryons on Satellite Kinematics}
\shortauthors{Baggen, van den Bosch and Mitra}

\author{Josephine F.W. Baggen$^1$\orcidlink{0009-0005-2295-7246}}
\author{Frank C. van den Bosch$^1$\orcidlink{0000-0003-3236-2068}}
\author{Kaustav Mitra$^{1,2}$\orcidlink{0000-0001-8073-4554}}

\affiliation{$^1$Department of Astronomy, Yale University, PO. Box 208101, New Haven, CT 06520-8101}
\affiliation{$^2$HEP Division, Argonne National Laboratory, 9700 South Cass Avenue, Lemont, IL 60439, USA}

\email{josephine.baggen@yale.edu}

\label{firstpage}

%%%%%%%%%%%%%%%%%%%%%%%%%%%%%%%%%%%%%%%%%%%%%%%%%%%%%%%%%%%

\begin{abstract}
   Satellite kinematics offers a powerful method to infer dynamical halo masses and has been demonstrated to yield tight constraints on the galaxy–halo connection. However, previous studies have assumed that the halos in which the satellites orbit are composed solely of dark matter, neglecting the role of baryons. Here, we develop an analytical model incorporating stars, gas, and the adiabatic response of dark matter to assess the impact of baryonic effects on the inference from satellite kinematics. The model covers halos in the mass range $10^{12}-10^{15}\Msun$ and is tuned to agree with well-established observational scaling relations. In addition, the model uses simple functional forms for the mass fractions of ejected baryons and diffuse halo stars, calibrated to the median trends in the EAGLE hydrodynamical simulations. We find that baryonic effects mainly result in a reduction of the satellite line-of-sight velocity dispersion due to the ejection of baryons and the resulting response of the dark matter halo. The effect is minimal (less than $1\%$) for the most massive halos, but reaches $\sim 5-6\%$ for halos in the mass range $10^{12} - 10^{13} \Msun$, and up to $8\%$ in extreme cases. We propose a simple formalism for correcting the satellite line-of-sight velocity dispersion for baryonic effects, and for marginalizing over the uncertainties. We integrate this correction function into \Basilisc, a Bayesian hierarchical inference method applied to satellite kinematics data extracted from large redshift surveys, and find that this shifts central galaxies to higher inferred halo masses at fixed luminosity by up to $\sim$0.3 dex. In a forthcoming work, we demonstrate that these few-percent level baryonic effects can have a non-negligible impact on the inference of cosmological parameters, motivating a novel approach to constraining the efficiency of feedback processes associated with galaxy formation.
\end{abstract} 

%%%%%%%%%%%%%%%%%%%%%%%%%%%%%%%%%%%%%%%%%%%%%%%%%%%%%%%%%


\maketitle

%%%%%%%%%%%%%%%%%%%%%%%%%%%%%%%%%%%%%%%%%%%%%%%%%%%%%%%%

\section{Introduction}
\label{sec:intro}

In the $\Lambda$CDM framework, which serves as the well-established cosmological model, galaxies form and reside in the centers of dark matter halos \citep[e.g.][and references therein]{MBW2010}. The link between the central galaxy and its host halo, known as the galaxy-halo connection, is fundamental for both galaxy formation physics and cosmology. On the one hand, it reflects the final outcome of the complex non-linear processes of galaxy formation that cannot be solved analytically, making it a crucial tool for testing and tuning semi-analytical models \citep[e.g.][]{Somerville2008} and hydrodynamical simulations \citep[e.g.][]{crain2015, Genel2014}. On the other hand, it links the observable light emitted by galaxies to the underlying dark matter, and thus, understanding the galaxy-halo connection is required when one wants to constrain cosmological parameters with galaxy surveys. 

The advent of wide-field galaxy surveys has enabled data-driven approaches to empirically infer the galaxy-halo connection and its evolution over time. Popular methods include galaxy clustering  \citep[e.g.][]{BerlindWeinberg2002,YangMoBosch2003, vandenBosch2007, Zehavi2011, HearinWatson2013, Guo2015a, Guo2015b, Guo2016, Zentner2019},  galaxy–galaxy lensing \citep[e.g.][]{GuzikSeljak2002, Mandelbaum2006, Mandelbaum2016, Sonnenfeld2018, Leauthaud2012, Leauthaud2017}, 
and subhalo abundance matching (SHAM) \citep{Marinoni_Hudson_2002, ConroyWechslerKravtsov2006, Vale_Ostriker2004, Reddick2013}. We refer to \citet{Wechsler2018} for a comprehensive review of the different approaches to probe and understand the galaxy-halo connection. 

A less commonly used technique for probing the galaxy-halo connection is satellite kinematics. This method represents the earliest approach, dating back to the first evidence for dark matter by \citet{Zwicky1933}. It involves measuring the line-of-sight velocities of satellite galaxies relative to their central galaxy. Since satellite galaxies are assumed to trace the dark matter potential wells, their dynamics can be used to constrain the gravitational potential and mass of the host halo.  Although some concerns have been raised about this method \citep[for an extensive review, see][and references therein]{Lange2019a}, numerous studies have progressively refined satellite kinematics and demonstrated its utility as a robust tool for constraining the galaxy–halo connection \citep{vandenBosch2004, Conroy2007,More2009I, More2011, WojtakMamon2013,Lange2019a, Lange2019b, Mitra2024, vandenBosch2019}. In particular, \Basilisc, a Bayesian hierarchical inference formalism, introduced by \citet*{vandenBosch2019} and refined by \citet{Mitra2024}, has demonstrated exceptional power to probe the galaxy-halo connection using the kinematics of satellite galaxies in large redshift surveys such as the Sloan Digital Sky Survey \citep[SDSS;][]{York2000}. \Basilisk forward models the 2-D projected phase-space distribution of satellite and central galaxy pairs. A key advantage of this forward-modelling approach is its ability to directly compute the likelihood of the full projected phase-space data given the model without the need for stacking the data in luminosity bins. Consequently, it can simultaneously solve for halo mass, the radial profile of the satellite galaxies and orbital anisotropy of the satellite galaxies, while properly accounting for scatter in the galaxy–halo connection. In addition, it properly accounts for biases and selection effects in the data, such as incompleteness, fibre collisions, and interlopers \citep{Lange2019a,vandenBosch2019, Mitra2024}. Recently, \citet{Mitra2024} applied \Basilisk to SDSS-DR7 data, yielding inferences on the galaxy–halo connection that are consistent with, yet tighter than, previous constraints derived from other methods like galaxy clustering and galaxy–galaxy lensing, while also overcoming the mass-anisotropy degeneracy and without being sensitive to halo assembly bias.

An important assumption in satellite kinematics studies is the halo's density profile, which shapes the gravitational potential. Often, such studies, including those conducted with \Basilisc, assume that halos are composed solely of dark matter, neglecting the presence of baryons. However, baryons, in the form of gas and stars, can play a critical role in shaping the structure and dynamics of dark matter halos. Processes such as gas cooling, star formation, and feedback from supernovae or active galactic nuclei (AGN) can modify the distribution of both baryons and dark matter.  Early models of baryonic effects suggested that gas cooling and dissipation, followed by star formation, can deepen the gravitational potential well \citep{WhiteRees1978, WhiteFrenk1991} and lead to adiabatic contraction of dark matter \citep*{Blumenthal1986}. With the advent of cosmological hydrodynamical simulations,  detailed studies have further examined the role of baryons in shaping halo structure, clustering, and abundance \citep[e.g.][]{Duffy2010, Schaller2015a,Schaller2015b,Sorini2024, Beltz-MohrmannBerlind2021, Castro2024}. These effects are highly complex, as the influence of baryonic physics depends on factors such as halo mass, environment, and redshift. Moreover, the impact of baryons is extremely sensitive to the specific feedback prescriptions, which differ significantly across hydrodynamical simulations \citep[see review by][]{Sommerville2015}. Despite these challenges, hydrodynamical simulations are often used for correcting for ``baryonic effects", that is deviations from dark-matter-only models, when constraining for example the matter power spectrum \citep[e.g.][]{White2004, vanDaalen2011, Schaller2025, Schaye2023, Hernandez-Aguayo2023} and weak lensing observables  \citep{Broxterman2024, Sembolini2011}. 

An alternative approach is to quantify baryonic effects analytically, providing a flexible tool to complement hydrodynamical simulations and systematically test their impact on cosmological and large-scale structure inferences. While such an approach has been used to study baryonic effects on the matter power spectrum and weak lensing \citep{Fedeli2014a, Fedeli2014b, Sembolini2011, Mead2015, Schneider2015, Schneider2019}, to our knowledge baryonic corrections have not yet been applied to the inference from satellite kinematics. The main goal of this paper is to rectify this situation and to study how baryons impact the line-of-sight velocity dispersion of satellite galaxies modelled as tracers of the underlying gravitational potential well. We do so by comparing the results of a dark-matter-only (DMO) model with one in which the halo hosts a central galaxy and an extended distribution of diffuse stars and gas. Crucially, we examine how the ejection of baryons from the host halo due to feedback processes impacts the potential well, and consequently, the satellite kinematics. We use these results to develop correction functions that can be used to correct the line-of-sight velocity dispersions of satellites computed using a DMO model for the presence of baryons. We also explore conservative bounds on this baryonic correction model that can be used to marginalize over uncertainties arising from incomplete understanding of physical processes associated with galaxy formation. We specifically tailor these results so they can be used in \Basilisk when applied to SDSS data. In a companion paper \citep{Mitra2025arXiv251214889M_Basilisk_no_tension}, it is shown that these baryonic corrections, which are typically only at the level of a few percent, can have an appreciable impact on the inference, affecting both the inferred galaxy halo connection as well as cosmological constraints. 

The structure of this paper is as follows. In Section~\ref{sec:methodology}, we describe our analytical approach for modelling the line-of-sight velocity dispersion profiles of satellites in a halo largely following the methodology developed for \Basilisc. In Section~\ref{sec:eagle}, we calibrate and validate the model against data from the hydrodynamical EAGLE ('Evolution and Assembly of GaLaxies and their Environment') simulations \citep{schaye2015, crain2015}. In Section~\ref{sec:results}, we use the model to assess how baryonic effects impact the line-of-sight velocity dispersion of satellite galaxies, and we propose a simple `baryonification' method to correct satellite kinematics modelling for the presence of baryons. Finally, Section~\ref{sec:fibre} discusses the fortunate impact of fibre collisions in spectroscopic survey data, and Section~\ref{sec:discussion} summarizes our main findings.

Throughout this paper, we adopt the flat Planck18 CDM cosmology \citep[with BAO constraints, Table 2 in][]{Planck2020}.
%, with $H_0$=67.66 km s$^{-1}$ Mpc$^{-1}$, $\Omega_{\rm m,0}$ = 0.3111,   $\Omega_{\rm b,0}$ = 0.0490, $n_{\rm s}=0.9665$ and $\sigma_8 = 0.810$. 

\section{Methodology}
\label{sec:methodology}

\subsection{Satellite kinematics}
\label{sec:methodology:satellitekinematics}

The main observable considered in this study is the line-of-sight velocity dispersion of satellite galaxies as measured over some aperture, $\sigma_{\rm ap}$. Throughout we assume halos (with or without baryons) to be spherically symmetric. In addition, we assume that satellite galaxies can be treated as a virialized tracer population of the underlying potential, $\Phi(r)$. This implies that their radial velocity dispersion at halo-centric distance $r$ can be written as
\begin{align}
\label{eq:methodology:sigmar}
 \sigma_\rmr^2(r) = \frac{G}{r^{2\beta} \, n_{\mathrm{sat}}(r)} \int_r^{\infty} r'^{2\beta - 2} n_{\mathrm{sat}}(r') \, M_{\rm tot}(r) \, \mathrm{d}r'\,,    
\end{align}
\citep[e.g.][]{vandenBosch2004, LokasMamon2003MNRAS.343..401L}, where $M_{\rm tot}(r)$ is the total enclosed mass profile, $n_{\mathrm{sat}}(r)$ is the radial number density profile of the satellites, and 
\begin{equation}
 \beta = 1 -\frac{\sigma_\theta^2 + \sigma_\phi^2}{2\sigma_\rmr^2}\,,
\end{equation}
is the velocity anisotropy parameter relating the velocity dispersions in the tangential and radial directions \citep{Binney1980}. Unless stated otherwise, we set $\beta=0$ implying isotropy. 

Integrating the velocity dispersion in the line-of-sight direction along the line-of-sight, weighted by the number density of the tracer population, yields the line-of-sight velocity dispersion \citep[e.g.][]{BinneyMamon1982}
\begin{equation}
\label{eq:methodology:sigmalos}
\sigma_{\mathrm{los}}^2(R_\rmp) = \frac{2}{\Sigma(R_{\rm p})} \int_{R_{\rm p}}^{\infty} \left[1-\beta\frac{R_\rmp^2}{r^2}\right] \, n_{\mathrm{sat}}(r) \, \sigma_\rmr^2(r) \frac{r \,\mathrm{d}r}{\sqrt{r^2 -R_\rmp^2}}\,.
\end{equation}
Here $R_\rmp$ is the projected separation from the center of the halo and
\begin{equation}
\label{eq:methodology:SigmaRp}
\Sigma(R_\rmp) = 2 \int_{R_{\rm p}}^{\infty} n_{\mathrm{sat}}(r) \, \frac{r \,\mathrm{dr}}{\sqrt{r^2-R_\rmp^2}}\,,
\end{equation}
is the projected number density distribution of satellite galaxies. These can be used to compute the line-of-sight velocity dispersion measured (or integrated) over an annular aperture with $R_{\rm min} < R_\rmp < R_{\rm max}$ as
\begin{equation}
\label{eq:methodology:sigmaap}
\sigma_{\mathrm{ap}} =\frac{2\pi\int_{R_\mathrm{min}}^{R_\mathrm{max}} \Sigma(R_\rmp) \,\sigma_{\mathrm{los}}(R_\rmp) \, R_\rmp \, \mathrm{d}R_\rmp}{2\pi\int_{R_\mathrm{min}}^{R_\mathrm{max}} \Sigma(R_\rmp) \, R_\rmp \, \mathrm{d}R_\rmp}\,.    
\end{equation}

Throughout we follow \citet{Mitra2024} and assume that the radial number density distribution of satellite galaxies follow a generalized \citet*{NavarroFrenkWhite1996} (NFW) profile given by
\begin{equation}\label{nsat}
 n_{\mathrm{sat}}(r) \propto \left(\frac{r}{\calR \, r_\rms}\right)^{-\gamma} \left(1+\frac{r}{\calR \, r_\rms}\right)^{\gamma-3}\,,
\end{equation}
\citep[see also][]{More2009II, Guo2012, Cacciato2013, vandenBosch2019, Lange2019b}. Here $r_\rms$ is the scale radius of the dark matter, while $\calR$ and $\gamma$ are free parameters that quantify how the radial profile of satellite galaxies differs from that of dark matter particles. If $\gamma=\mathcal{R} = 1$, satellites are unbiased tracers of the underlying dark matter distribution, following the same NFW profile as the dark matter. For our fiducial model, we adopt $\gamma=1$ and $\mathcal{R}=2$, which roughly matches the radial profile inferred for satellite galaxies in the SDSS data ($\gamma=0.94$ and $\mathcal{R}=1.7$) \citep[][]{Lange2019b, Mitra2024}. Throughout we also follow \citet{Mitra2024} in assuming that $n_{\rm sat}(r)$ extends out to a `splashback' radius of $r_{\rm sp} = 2 r_{\rm vir}$, %with $r_{\rm vir}$ the halo virial radius inside of which the average density is 97 times the critical density \citep{BryanNorman1998}. 
with $r_{\rm vir}$ the halo virial radius defined using the redshift-dependent virial overdensity with respect to the critical density following \citet{BryanNorman1998}.
For $r > r_{\rm sp}$ we set $n_{\rm sat}(r) = 0$, though in practice we simply set the upper integration limits in equations~(\ref{eq:methodology:sigmalos}) and~(\ref{eq:methodology:SigmaRp}) to $r_{\rm sp}$.

We have experimented with several different values for $\gamma$, $\calR$, and $\beta$. Although these parameters significantly impact $\sigma_{\rm los}(R_\rmp)$, and thus $\sigma_{\rm ap}$, they have a  negligible impact on the {\it ratio} of $\sigma_{\rm ap}$ values for the cases with and without baryons (see Appendix~\ref{sec:appendix:negligible_parameters}). Hence, for the purpose of exploring how baryons impact $\sigma_{\rm ap}$ we will restrict ourselves to our fiducial values for $\gamma$, $\mathcal{R}$, and $\beta$ without loss of generality.

Finally, we emphasize that the primary observable considered here is the line-of-sight velocity dispersion, i.e. the second moment of the line-of-sight velocity distribution (LOSVD). However, higher-order moments of the LOSVD, such as the skewness and kurtosis, also contain valuable information regarding the kinematics \citep[e.g.,][]{vdMarel.Franx.93}. Indeed, \Basilisk uses the kurtosis in its analysis, which helps to break the mass-anisotropy degeneracy \citep[][]{BinneyMamon1982}. However, as we demonstrate in Appendix~\ref{sec:appendix:kurtosis}, baryons have a negligible effect on the kurtosis over the radial range of interest, which is why in what follows, we focus solely on the line-of-sight velocity dispersion.

%\textbf{While our primary observable is the line-of-sight velocity dispersion, i.e. the second velocity moment, we also examine the fourth velocity moment and the effects of baryons on the resulting kurtosis profiles, which we present in Appendix~\ref{sec:appendix:kurtosis}.}

%
\begin{table*}
\centering
\begin{tabular}{p{0.08\linewidth}p{0.08\linewidth}p{0.44\linewidth}p{0.31\linewidth}}
\hline
\noalign{\vskip 2pt}
 & Parameter & Description & Fiducial \\ 
 \hline
 \noalign{\vskip 2pt}
Initial setup & $M_{\mathrm{DMO}}$& Total mass of the halo in the DMO model (defined as $M_{\mathrm{200c,DMO}}$). \\
 & $c_{200}$ & Concentration of the halo derived from concentration-halo mass relation. & \citet{DiemerJoyce2019}  \\
 & $z$ & Redshift of the halo. & $z=0.1$ \\
  & $\mathcal{R}$, $\gamma$ &  Parameters describing the radial profile of the satellites. & $\mathcal{R}=2$, $\gamma=1$  \\
 & $\beta$ &  Velocity anisotropy parameter. & $\beta=0$ \\  
  & $f_\mathrm{DM}$  &  Dark matter fraction inside $r_{200}$. & 
  $1.0$ for DMO, $0.84$ with baryons \\
& $R_\mathrm{min}$  &  Minimum aperture radius. & 55$\arcsec$ (SDSS)  \\
& $R_\mathrm{max}$  & Maximum aperture radius. & 0.375$r_{\mathrm{vir}}$  \\
\hline
\noalign{\vskip 2pt}
Model baryons & $f_\mathrm{*, cen}$ & Mass fraction of central galaxy, derived from the stellar-halo mass relation. 
& \citet{Moster2013} \\
 & a & Scale radius of central galaxy follows from the size-stellar mass relation. 
 & \citet{Shen2003} \\
  & $f_\mathrm{*, diffuse}$ & Fraction of mass in the diffuse stellar component. & \makecell[lt]{Best-fit derived from EAGLE: \\ $f_{\mathrm{\ast, diffuse, max}}=0.06$, $\log M_{\mathrm{char, diffuse}}=12.8$}
   \\
   & $\eta$ & Concentration of diffuse stellar component, defined as $c_{\rm *, diffuse} = \eta c_{200}$. & $\eta=3$ \\
 & $f_\mathrm{eject}$ & Mass fraction ejected from halo. &  \makecell[lt]{Best-fit derived from EAGLE: \\ $f_{\mathrm{ eject, max}}=0.66$, $\log M_{\mathrm{char, eject}}=13.2$, $\alpha=1.8$}
  \\
 & $x_\mathrm{tr}$ & Radius where gas density profile transitions from polytropic to NFW. & $x_\mathrm{tr}=c_{200}/\sqrt{5}$
  \\ 
 & $\nu$ & Halo response (equation~\ref{eq:methodology:adiabatic_nu}).  
 & $\nu=0, \nu=1$ \\ 
 \hline
\end{tabular}
 \caption{Parameters of the analytical halo model. In the DMO model, only the parameters in the top rows   (initial setup) are used, whereas the model with baryons includes several additional parameters to describe the properties of stars, gas and halo response.}
\label{tab:methodology:halomodel}
\end{table*}

\subsection{Mass models}
\label{sec:methodology:massmodels}

The key ingredient for computing $\sigma_{\rm ap}$ is a model for the total enclosed mass, $M_{\rm tot}(r)$, which enters the expression for the radial velocity dispersion given by equation~(\ref{eq:methodology:sigmar}). Here we describe the two mass models that we use for our investigation: a DMO model and a baryonic model.  

Throughout we use $\MDMO$ to refer to the mass of the dark matter halo in the DMO case, which we define as the mass inside the radius $r_{200}$ within which the mean density is 200 times the critical density. For the baryon model, we assume a \textit{universal} baryon fraction $f_{\rm b} = \Omega_\rmb/\Omega_\rmm = 0.16$, and correspondingly define the dark matter fraction as $f_{\rm DM} = 1 - f_{\rm b} = 0.84$, such that the dark matter mass is $M_{\rm DM} = f_{\rm DM} \MDMO$. In addition to the dark matter, the baryon model includes a central galaxy with stellar mass $M_{\mathrm{\ast, cen}}$, a diffuse stellar component with mass $M_{\mathrm{\ast, diffuse}}$ and gas component with mass $M_{\mathrm{gas}}$. We also introduce an ejected baryonic component, associated with galaxy formation-driven feedback processes, with a mass given by
\begin{equation}
\label{eq:methodology:feject}
M_{\mathrm{eject}} = f_{\rm b} \MDMO - M_{\mathrm{\ast, cen}} - M_{\mathrm{\ast, diffuse}} - M_{\mathrm{gas}}\,.
\end{equation}

For convenience, we define the following baryonic mass fractions:
\begin{center}
\begin{align}
\hspace{1cm}f_{\ast,\mathrm{cen}} &= \frac{M_{\ast,\mathrm{cen}}}{M_{\mathrm{b}}} \,, &
f_{\ast,\mathrm{diffuse}} &= \frac{M_{\ast,\mathrm{diffuse}}}{M_{\mathrm{b}}} \,, \nonumber \\
\hspace{1cm} f_{\mathrm{gas}} &= \frac{M_{\mathrm{gas}}}{M_{\mathrm{b}}} \,, &
f_{\mathrm{eject}} &= \frac{M_{\mathrm{eject}}}{M_{\mathrm{b}}} \,.
\label{eq:baryon_fractions}
\end{align}
\end{center}
where $M_{\rm b} =f_{\rm b} \MDMO$ represents the total baryonic mass associated with the halo under the assumption of a universal baryon fraction. By construction, these mass fractions satisfy $f_{\mathrm{\ast, cen}} + f_{\mathrm{\ast, diffuse}} + f_{\mathrm{gas}} + f_{\mathrm{eject}} = 1$.

Throughout, we assume that the ejected baryonic component has been fully removed from the halo (i.e., expelled beyond the splashback radius) and therefore no longer contributes to the halo's gravitational potential. Hence, we have that the total enclosed mass profile for the baryon model can be written as

\begin{align}
 M_{\rm tot}(r) = & M_{\rm DM}(r) + M_{\rm \ast, cen}(r) \, \notag \\
 & + M_{\rm \ast, diffuse}(r) + M_{\rm gas}(r)\,.
 \end{align}

We further define $M_{\rm tot} \equiv M_{\rm tot}(r_{200})$ 
%\st{as the total bound mass of the halo in the baryonic model, which can equivalently be expressed as} 
as the total bound mass of the halo in the baryonic model, which is related to the DMO mass according to 
\begin{equation} M_{\rm tot} = M_{\rm DMO}\left(1 - f_{\rm b}f_{\rm eject}\right)\,.
\end{equation} 

In what follows, we describe the modelling of each of these mass components in detail.
 
\subsubsection{Dark Matter}
\label{sec:methodology:darkmatter}

For both the DMO model and the baryonic model, we assume that the dark matter halo follows an NFW density profile:
\begin{equation}
\label{eq:methodology:nfwdensityprofile}
\rho_{\mathrm{NFW}}(r) = \rho_{\rm crit} {\delta_{\rm 200} \over (r/r_\rms) \, ( 1 + r/r_\rms)^{2}}\,.
\end{equation}
Here $r_\rms$ is a characteristic scale radius, $\rho_{\rm crit}$ is the critical density for closure, and $\delta_{200}$ is a characteristic overdensity given by
\begin{equation}
\delta_{200} = \frac{200}{3} \frac{c_{200}^3}{\ln(1+c_{200}) -c_{200}/(1+c_{200})}\,, 
\end{equation}
with $c_{200} = r_{200}/r_\rms$ the halo concentration. Throughout we compute this concentration parameter using the redshift-dependent concentration-mass relation of \citet{DiemerJoyce2019}.

The enclosed mass profile for the NFW profile is given by
\begin{equation}
\label{eq:methodology:NFWmassprofile}
M_{\rm DM}(r) = f_{\mathrm{DM}} \MDMO \, \frac{\mu(r/r_\rms)}{\mu(c_{200})}\,,
\end{equation}
where
\begin{equation}
\mu(x) = \ln(1+x) - \frac{x}{1+x}\,,   
\end{equation}
and $f_{\rm DM} =1$ $(0.84)$ for the DMO (baryonic) model. 

\subsubsection{Central Galaxy}
\label{sec:methodology:central}

The stellar component of the central galaxy is modeled as a spherical \citet{Hernquist199} profile,
\begin{equation}
\rho_{\mathrm{\ast, cen}}(r) = \frac{M_{\mathrm{\ast, cen}}}{2\pi} \frac{a}{r(r+a)^3}\,,    
\end{equation}
which implies
\begin{equation}
\label{eq:methodology:hernquistmassprofile}
M_{\mathrm{\ast, cen}}(r) = M_{\mathrm{\ast, cen}} \frac{r^2}{(r+a)^2}\,. 
\end{equation}

The stellar mass of the central galaxy, $M_{\mathrm{\ast, cen}}$, is taken from the empirical stellar mass-halo mass relation (SHMR) of \citet{Moster2013}, which is in good agreement with that of numerous other studies \citep[e.g.,][]{Guo2010, More2011, Yang2012, Behroozi2013, Kravtsov2018, Yang2012}. Since the SHMR of \citet{Moster2013} is obtained using abundance matching based on dark matter halos extracted from a DMO simulation we interpret it as a relation of the form  $M_{\mathrm{\ast, cen}}(\MDMO)$; i.e., we infer the stellar mass of the central from the halo mass in the DMO case, not from the dark matter mass in the baryonic model.

In order to set the scale radius, $a$, for the Hernquist profile, we first determine the effective radius, $R_\rme$, from the size-mass relation for early-type galaxies of \citet{Shen2003}, who found that $\log(R_{\rm e}/\kpc) = 0.56 \log(M_\ast/\Msun) -5.54$. 
We then convert the effective radius to the corresponding Hernquist scale radius using $a = 0.55 \,R_\rme$ \citep[][]{Hernquist199}.

\subsubsection{Diffuse stellar component}
\label{sec:methodology:reststellar}

The stellar mass budget of a halo typically includes three components: a central galaxy, stars in satellites, and a stellar halo. Here we model the combination of the latter two as a single component, to which we refer as the diffuse stellar component, and which thus includes all stars within the confines of the halo that are not part of the central galaxy. 

Whereas $f_{\mathrm{\ast, cen}}(\MDMO)$ is tightly constrained via the SHMR, the baryonic mass fraction of the diffuse stellar component, $f_{\mathrm{\ast, diffuse}}$, is poorly constrained. The main reason is that stellar halos typically have extremely low surface brightness, making them difficult to detect. In addition, the outer light profile of the central galaxy typically blends smoothly into that of the stellar halo, hampering a unique decomposition, both in observations \citep[e.g.,][]{Huang2018} and in numerical simulations \citep[e.g.,][]{Pillepich2018}. Despite these challenges, it has become clear that $f_{\mathrm{\ast, diffuse}}$ increases with halo mass, albeit with considerable scatter \citep[][]{Merritt2016, Deason2019, Ragusa2023, Montes2019}. We describe this mass dependence using the following simple analytical form:
\begin{equation}
\label{eq:methodology:fdiffuse}
f_{\mathrm{\ast, diffuse}}(\MDMO) = \frac{f_{\mathrm{\ast, diffuse, max}}}{2}\left[1+ \erf\left(\log\frac{\MDMO}{M_{\mathrm{char, diffuse}}}\right)\right]\,,
\end{equation}
which transitions from $f_{\mathrm{\ast, diffuse}} = 0$ for $\MDMO \ll M_{\rm char, diffuse}$ to a maximum of $f_{\mathrm{\ast, diffuse, max}}$ when $\MDMO \gg M_{\rm char, diffuse}$. We treat $f_{\mathrm{\ast, diffuse, max}}$ and $M_{\rm char, diffuse}$ as free parameters that we tune to data from the EAGLE simulation for our fiducial model (see Section~\ref{sec:eagle}).

We assume that the diffuse stellar component, which describes both the stellar halo as well as the stellar mass of all the satellite galaxies, roughly follows the density distribution of the dark matter. As already mentioned in Section~\ref{sec:methodology:satellitekinematics}, this is indeed a good approximation for the radial distribution of satellite galaxies. In addition, both observations \citep[e.g.,][]{Montes2019} and hydrodynamical simulations \citep[e.g.,][]{Yoo2024, ContrerasSantos2024, AlonsoAsensio2020} have indicated that the stars that make up the stellar halo are a good tracer of the dark matter. We therefore model the density profile of the diffuse stellar component as an NFW profile with a concentration parameter $c_{\mathrm{*, diffuse}} = \eta c_{200}$. For our fiducial model we set $\eta=3$, which yields a good fit to the distribution of diffuse light (ICL) in groups and clusters \citep[][]{Contini2020}. Note that this approach ignores the fact that satellite galaxies contribute localized density peaks. Rather, the mass of the satellites is modeled as being distributed smoothly following a spherically symmetric profile. As long as the stellar masses of individual satellite galaxies are sufficiently small compared to the host halo, this oversimplification will not have a significant impact on our results. Recall anyways that we assume throughout that satellite galaxies are a kinematic tracer population of the host halo's gravitational potential.

\subsubsection{Gas}
\label{sec:methodology:gas}

In order to model the baryonic mass fraction of gas that resides within a halo characterized by DMO mass $\MDMO$, we proceed as follows. We characterize the mass dependence of the {\it ejected} baryonic mass fraction, $f_{\rm ejected}(\MDMO)$, from which we obtain the {\it bound} gas mass fraction following
\begin{equation}
\label{eq:methodology:fgas}
f_{\mathrm{gas}} = 1 - f_{\mathrm{\ast, cen}} - f_{\mathrm{\ast, diffuse}} - f_{\rm eject}\,. 
\end{equation}

Models of galaxy formation generally predict that more massive halos, with their deeper potential wells, are more effective at retaining baryons \citep[][]{Dekel.Silk.86}, a trend that is also supported by observations \citep[e.g.,][]{Dai2010}.  We therefore model the ejected baryon fraction, $f_{\rm eject}$, as a decreasing function of halo mass, using a functional form:
\begin{equation}
\label{eq:methodology:feject_function}
f_{\mathrm{eject}}(\MDMO) = \frac{f_{\mathrm{eject, max}}}{2}\left[1 - \erf\left(\alpha \, \mathrm{log}\frac{\MDMO}{M_{\mathrm{char, eject}}}\right)\right]\,.
\end{equation}
Hence, the ejected mass fraction transitions from a maximum of $f_{\mathrm{eject, max}}$ at the low mass end, to $f_{\rm eject} = 0$ at the high mass end. The steepness of the transition is controlled by $\alpha$, which we treat as a (positive) free parameter. The characteristic mass $M_{\rm char, eject}$ sets the DMO halo mass where $f_{\rm eject} = 0.5 f_{\rm eject, max}$. The bound gas fraction then automatically follows from equation~(\ref{eq:methodology:fgas}). 

We model the bound gas component in halos using a single, smooth, spherically symmetric distribution.  In reality, the gas inside dark matter halos consists of multiple components, including the interstellar media (ISM) of the central and satellite galaxies, as well as the circumgalactic medium (CGM)\footnote{In the case of massive clusters, the CGM is typically referred to as the intracluster medium (ICM).}. However, we focus exclusively on the CGM/ICM component for two main reasons. First, the ISM contributes only a small fraction of the total gas mass \citep{TumilsonPeeplesWerk2017}. Second, due to the fibre collision scale (see Section~\ref{sec:fibre}), our measurements are insensitive to the inner halo regions where the ISM of the centrals resides. Instead, we primarily probe the diffuse CGM, which extends on scales of order the virial radius. The CGM itself is expected to be multiphase, consisting of cold ($T < 10^4$ K), cool ($T \sim 10^4$–$10^5$ K), warm ($T \sim 10^5$ K) and hot ($T \geq 10^6$ K) gas, giving rise to complex features such as clumps, filaments, and streams. Although the hot phase is believed to dominate the CGM mass budget in Milky Way–mass halos, this remains observationally uncertain \citep{TumilsonPeeplesWerk2017}. In contrast, at the mass scale of galaxy groups and clusters, X-ray observations provide unambiguous evidence that the hot ($T \sim 10^7$–$10^8$ K) ICM is the dominant component.

We assume the gas is in hydrostatic equilibrium within the gravitational potential of the dark matter halo and follows a polytropic equation of state, $P \propto \rho^{\Gamma}$. Under these assumptions, the gas density profile can be expressed as \citep[e.g.][]{Komatsu.Seljak.2001, Suto.Sasaki.Makino.1998, Martizzi.Teyssier.Moore.2013}: 
\begin{equation}
\label{eq:methodology:gasprofile}
    \rho_{\rm gas}(x) = \rho_0 \left[\frac{\ln(1+x)}{x}\right]^{1/(\Gamma -1)}\,,
\end{equation}
where $x=r/r_{\rm s}$ and $\Gamma$ is the polytropic index. We require that the slope of the gas density profile matches that of the dark matter at a certain transition radius $r_{\rm tr} = x_{\rm tr} r_{\rm s}$, following \citet{Komatsu.Seljak.2001}. This then fixes the value for $\Gamma$:
\begin{equation}
    \Gamma = 1+ \frac{(1+x_{\rm tr})\ln(1+x_{\rm tr})-x_{\rm tr}}{(1+3x_{\rm tr})\ln(1+x_{\rm tr})}\,.
\end{equation}

The gas profile is defined such that the gas follows equation~(\ref{eq:methodology:gasprofile}) for $x < x_{\rm tr}$ and transitions to an NFW profile for $x > x_{\rm tr}$. By construction, the transition is smooth, since the slopes of the two profiles are matched at $x = x_{\rm tr}$.  This composite form naturally yields a flat core in the central gas density and follows the dark matter distribution at larger radii. In the fiducial model, we adopt a transition at $x_{\rm tr}=c_{200}/\sqrt5$, corresponding to $r_{\rm tr}=r_{200}/\sqrt5$, following the approach of \citet{Schneider2015}. This choice is motivated by observational evidence that the transition from a flattened core to a steeper, NFW-like slope typically occurs at $r$=0.3-0.5$r_{200}$, as inferred from stacked X-ray and Sunyaev-Zel'dovich (SZ) measurements of nearby clusters \citep[e.g.][]{Makino1998, Suto.Sasaki.Makino.1998,Vikhlinin.etal.2006, Eckert2013, Ghirardini.etal.2019B}. It has also been shown to yield good agreement with gas distributions in hydrodynamical simulations of galaxy clusters \citep{Mohammed.etal.2014}.
%\citep[see also][]{Battaglia.etal.2012, Mohammed.etal.2014, Sorini2022}.

Given the limited observational constraints on the structure of the gas in lower-mass halos, we assume that this physically motivated and self-similar description for gas in galaxy clusters remains valid across all halo mass scales. We compare our analytic gas profiles with those extracted from EAGLE halo stacks centered at $M_{\rm DMO} = 10^{12.5} \Msun$ and $10^{13.5} \Msun$, finding generally good agreement (see Section~\ref{sec:results:comparisoneagle}). At these mass scales, adopting a larger transition radius \citep[e.g. $x_{\rm tr} = c_{200}$ following ][]{Komatsu.Seljak.2001} yields a slightly better match to the simulated gas density profiles but has a negligible effect on the velocity dispersion at these mass scales. At higher halo masses ($M_{\rm DMO} \sim 10^{15} \Msun$), this alternative choice impacts the velocity dispersion with a decrease in velocity dispersion at the $0.5\%$ level (see Appendix~\ref{sec:appendix:negligible_parameters}). We account for this effect in Section~\ref{sec:results:variations_from_fiducial} where we explore extreme baryon models. We note that the polytropic index is largely insensitive to the choice of $x_{\rm tr}$, remaining at $\Gamma \approx 1.2$ across concentrations $4 < c_{200} < 7$, consistent with previous analytical and simulation results \citep{Komatsu.Seljak.2001, Ascasibar.etal.2006, Ghirardini.etal2017}, 

The enclosed mass profile for the gas component, $M_{\rm gas}(r)$ is computed by numerically integrating this density profile. The normalization $\rho_0$ is set such that the total gas mass fraction within the halo corresponds to the desired value of $f_{\rm gas}$.
\begin{figure*}
    \centering
\includegraphics[width=\linewidth, trim={0cm 0.2cm 0cm 0cm}, clip]{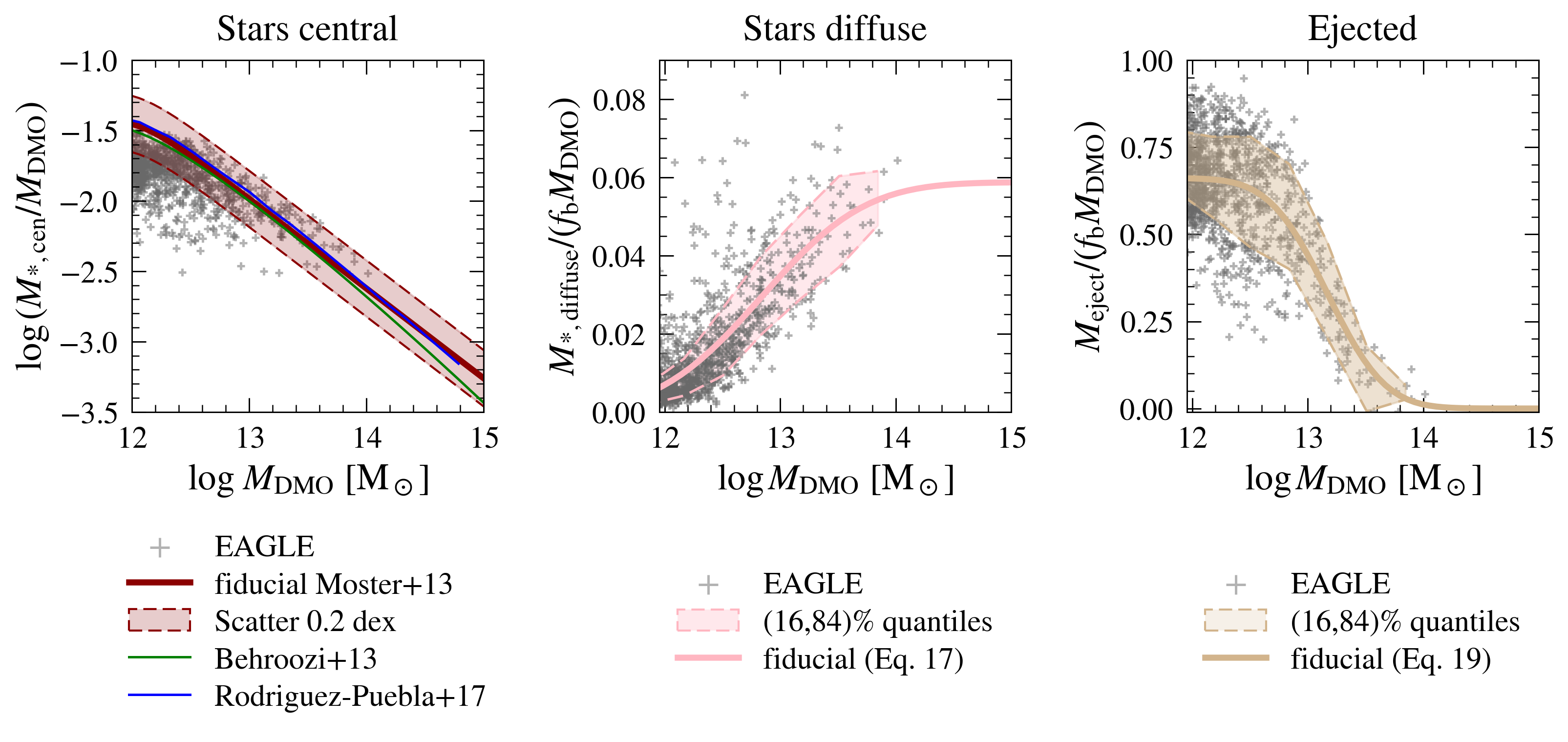}
    \caption{\textit{Left:} Fraction of stellar mass in the central galaxy relative to the total DMO halo mass. The fiducial model follows the stellar–halo mass relation of \citet{Moster2013}.  For comparison, we overplot the empirical SHMRs of \citet{Moster2013}, \citet{Behroozi2013} and \citet{RodriguezPuebla2017}, as indicated, which are in excellent agreement with each other. The red shaded band marks 0.2 dex scatter around the relation of \citet{Moster2013}, which, based on estimates of the scatter in the SHMR, indicates the band that is expected to enclose about 68 percent of all galaxies. The gray points indicate relaxed halos from the EAGLE simulation, which underpredict the observed relation at low halo masses.  \textit{Middle}: Fraction of stellar mass in the diffuse stellar component as a function of halo mass, defined as $f_{\mathrm{\ast, diffuse}} = M_{\rm \ast, diffuse}/(f_{\rm b}M_{\mathrm{DMO}})$, where $M_{\rm \ast, diffuse}$ is the total stellar mass within the virial radius minus the central galaxy mass.  The solid pink curve is a fit to the EAGLE data of the form of equation~(\ref{eq:methodology:fdiffuse}). \textit{Right}: Fraction of ejected baryonic mass as a function of halo mass, defined as $f_{\mathrm{eject}} = M_{\rm eject}/(f_{\rm b}M_{\mathrm{DMO}})$, where $M_{\rm eject}$ is determined through equation~(\ref{eq:methodology:feject}). The solid brown curve is a fit to the EAGLE data of the form of equation~(\ref{eq:methodology:fgas}). The shaded bands in the middle and right panels indicate the 16th–84th percentile ranges, computed in ten equally spaced bins in halo mass.}
\label{fig:methodology:eagle_icl_gas_fits}
\end{figure*}

\subsection{Adiabatic response}
\label{sec:methodology:adiabaticresponse}

When gas cools and accumulates at the center of its dark matter halo, the dark matter particles will respond to the change in the gravitational potential well, causing the dark matter halo to contract. Similarly, when gas is ejected from the halo due to feedback processes, the dark matter particles becomes less bound, causing the halo to expand. In general, this response of the dark matter is difficult to model in detail. However, if the processes responsible for the change in the overall gravitational potential are sufficiently slow, the response can be assumed to be adiabatic, which is analytically tractable. In particular, if we assume all dark matter to be on circular orbits, and that the system is spherically symmetric, then $r M(r)$, which is proportional to the square of the specific angular momentum, is an adiabatic invariant \citep*{Blumenthal1986}. Hence, under these conditions we have that 
\begin{equation}
\label{eq:methdology:adiabatic_invariant}
    r_\rmf \, M_\rmf(r_\rmf) = r_\rmi \, M_\rmi(r_\rmi)\,,
\end{equation}
which can be used to solve the final post-adiabatic-response radius, $r_\rmf$, given the initial pre-adiabatic-response radius $r_\rmi$. In the case of our baryonic model we have
\begin{equation}\label{eq:methodology:AC}
  r_\rmf \, \left[M_{\rm bar}(r_\rmf) +   M_{\rm DM,f}(r_\rmf) \right]  
= r_\rmi \, M_{\rm DM,i}(r_\rmi)\,,
\end{equation}
where  $M_{\mathrm{bar}}(r_\rmf)$ is the sum of the bound baryonic components (central galaxy, diffuse stars and gas). We iteratively solve this equation for $r_\rmf(r_\rmi)$, from which we then compute the enclosed mass profile of the adiabatically-responded dark matter halo.

This methodology is only valid in the limit where the changes to the gravitational potential are adiabatic. Unfortunately, certain galaxy formation processes, such as galactic outflows, are clearly not in this limit. Also, the assumption of circular orbits is a clear oversimplification, and \citet{Gnedin2004} has shown that accounting for the fact that dark matter particles are on eccentric, rather than circular obits, results in a weaker response. Furthermore, in the method described above we are assuming that only the dark matter responds. In reality, following the response of the dark matter, the hot gas will readjust its density and temperature profile to maintain hydrostatic equilibrium, something we do not take into account. In order to be able to account for these uncertainties, we follow \citet{Gnedin2004} and \citet{Dutton2007} and introduce some flexibility in modelling the response of the dark matter halo. Specifically, we use a modified value for $r_\rmf$ given by
\begin{equation}
\label{eq:methodology:adiabatic_nu}
    r'_\rmf = r_\rmi \, \Upsilon^{\nu}\,,
\end{equation}
where $\Upsilon = r_{\rm f}/r_{\rm i}$ is the contraction factor obtained by numerically solving equation~(\ref{eq:methodology:AC}), and $\nu$ is another free parameter of our model which sets the 'strength' of the response. Setting $\nu=0$ is equivalent to no response of the dark matter, while setting $\nu=1$ preserves adiabatic invariants as described by \citet*{Blumenthal1986}. In general, by varying $\nu$ we can examine how strongly our results are impacted by uncertainties and complications related to how the gravitational potential responds to the galaxy formation process.

\section{Model Calibration and Validation}
\label{sec:eagle}

In order to validate our baryonic model and to assure that its fiducial parameters are reasonable, we tune and compare the model against the cosmological EAGLE ('Evolution and Assembly of GaLaxies and their Environment') simulations \citep{schaye2015, crain2015}. In particular, using data from both the hydrodynamical and the corresponding DMO runs, we examine how the various baryonic mass fractions in EAGLE scale with the DMO halo mass, and how the baryons are distributed within their host halos. We restrict our analysis to the $z=0$ simulation outputs.

We closely follow \citet{Schaller2015a} who examined the impact of baryons in the EAGLE simulations by linking halos between the hydrodynamical and DMO runs. We use the density and enclosed mass profiles from their work, kindly provided to us in electronic form. We only use halos that are relaxed, defined as halos for which the center of mass is separated less than 0.07 times the virial radius from the center of the potential \citep[see][]{Schaller2015a}. We make this selection cut to be consistent with the assumption that satellites are a relaxed tracer population, but we have verified that including the non-relaxed halos does not impact any of our results. In what follows, we restrict our analysis to the 834 halos with $M_{\rm DMO} \geq 10^{12} \Msun$, which is roughly the mass range probed by \Basilisc. In addition, focusing on the most massive halos ensures that we have adequate spatial resolution to resolve the radial profiles. 

We follow \citet{Matthee2017} and define the stellar mass of the central galaxy as the stellar mass mass within 30 kpc, i.e., $M_{\mathrm{\ast, cen}} = M_{\ast}(30\kpc)$. The remaining stellar mass within $r_{200}$ (as measured in the DMO simulation) is assigned to the diffuse stellar component; $M_{\mathrm{\ast, diffuse}} = M_{\ast}(r_{200}) - M_{\mathrm{\ast, cen}}$. Similarly, the gas mass is defined as the {\it total} mass in gas inside $r_{200}$, and the ejected mass is computed using equation~(\ref{eq:methodology:feject}). 

%$M_{\mathrm{eject}} = f_{\rm b}\,M_{\mathrm{ DMO}}- M_{\mathrm{gas}} - M_{\mathrm{\ast, cen}} - M_{\mathrm{\ast, diffuse}}$.

The gray scatter points in Fig.~\ref{fig:methodology:eagle_icl_gas_fits} show the results thus obtained. The left-hand panel plots $M_{\mathrm{\ast, cen}}$ as a function of $\MDMO$. We compare our fiducial model to several empirical stellar–halo mass relations from the literature, which are in excellent agreement with each other. 
%For comparison, we overplot the empirical SHMRs of \citet{Moster2013}, \citet*{Behroozi2013} and \citet{RodriguezPuebla2017}, as indicated, The red shaded band marks 0.2 dex scatter around the relation of \citet{Moster2013}, which, based on estimates of the scatter in the SHMR, indicates the band that is expected to enclose about 68 percent of all galaxies. 
For $\MDMO \gta 10^{13} \Msun$ the central stellar masses in the EAGLE simulation are in good agreement with these empirical constraints. At lower halo mass, though, EAGLE underpredicts the average stellar mass of its centrals, and with a relatively large halo-to-halo variance. Since we want our baryonic model to be in close agreement with empirical constraints, we use the SHMR of \citet{Moster2013} to assign stellar masses for the centrals (see Section~\ref{sec:methodology:central}), rather than a fit to the simulation results of EAGLE. 

The middle panel of Fig.~\ref{fig:methodology:eagle_icl_gas_fits} plots the baryonic mass fraction of the diffuse stellar component, $f_{\mathrm{\ast, diffuse}}$ as a function of $\MDMO$. The pink-shaded region marks the 16 to 84 percentile range as inferred from EAGLE. Fitting equation~(\ref{eq:methodology:fdiffuse}) to this data, we obtain the relation indicated by the thick solid pink line, which has $f_{\mathrm{\ast, diffuse, max}} = 0.0588$ and $\log M_{\mathrm{char, diffuse}} = 12.84$ as best-fit parameters. It shows that the diffuse stellar component only comprises about one percent of the entire baryonic mass budget for halos with a DMO mass of $\sim 10^{12} \Msun$. This fraction steadily increases with halo mass, reaching about 6\% at the mass scale of $\sim 10^{15} \mathrm{M}_\odot$, which  exceeds the mass fraction of the central galaxy. Overall, these results are in broad agreement with observational constraints (see previous section). We therefore adopt this best-fit relation as our fiducial model for the mass fraction of the diffuse stellar component. We acknowledge that different simulations may make different predictions for $f_{\mathrm{\ast, diffuse}}(\MDMO)$, but as we demonstrate in Appendix~\ref{sec:appendix:negligible_parameters}, changing $f_{\mathrm{\ast, diffuse}}(\MDMO)$ within reasonable limits has no significant impact on our results. Hence, this particular functional form does not impact the conclusions in this work. 

Finally, the right-hand panel of Fig.~\ref{fig:methodology:eagle_icl_gas_fits} plots the ejected mass fraction of baryons, $f_{\rm eject}$, as a function of $\MDMO$. The brown shaded region marks the 16-84 percentile range in EAGLE, while the thick solid line indicates the best-fit relation of the form of equation~(\ref{eq:methodology:feject_function}), which has $(f_{\mathrm{eject, max}}, \log M_{\mathrm{char, eject}}, \alpha) = (0.661, 13.17, 1.79)$ as best-fit parameters. This is the $f_{\mathrm{eject}}(\MDMO)$ that we adopt for our fiducial baryonic model. Note how $f_{\rm eject} \rightarrow 0$ at the high-mass end, in agreement with the fact that the baryonic mass fractions in clusters are observed to be in agreement with the universal baryon fraction \citep[e.g.,][]{Gonzalez2007, Gonzalez2013, Dai2010, Andreon2010, Chiu2018, Morandi2015}. At $\MDMO \sim 10^{12}\Msun$, though, the EAGLE simulation predicts that, on average, close to 70 percent of all baryons have been ejected. Not only does the EAGLE simulation predict a large halo-to-halo variance at the low-mass end, different cosmological simulations often make predictions for $\langle f_{\rm eject} \rangle(\MDMO)$ that are substantially different from each other due to different (subgrid) implementations of the various feedback processes at play \citep[e.g.][]{Ayromlou.Nelson.Pillepich.2023,Wright.etal.2024}. In Section~\ref{sec:results:variations_from_fiducial} we therefore explore the impact of alternative $\langle f_{\rm eject} \rangle(\MDMO)$ relations by varying the parameters $f_{\rm eject, max}$, $\log(M_{\mathrm{char, eject}})$ and $\alpha$.

\section{Results}
\label{sec:results}

\begin{figure*}
    \centering
    \hspace{0.5cm}
    \includegraphics[width=\linewidth, trim = {0cm 3.5cm 0cm 0cm}, clip]{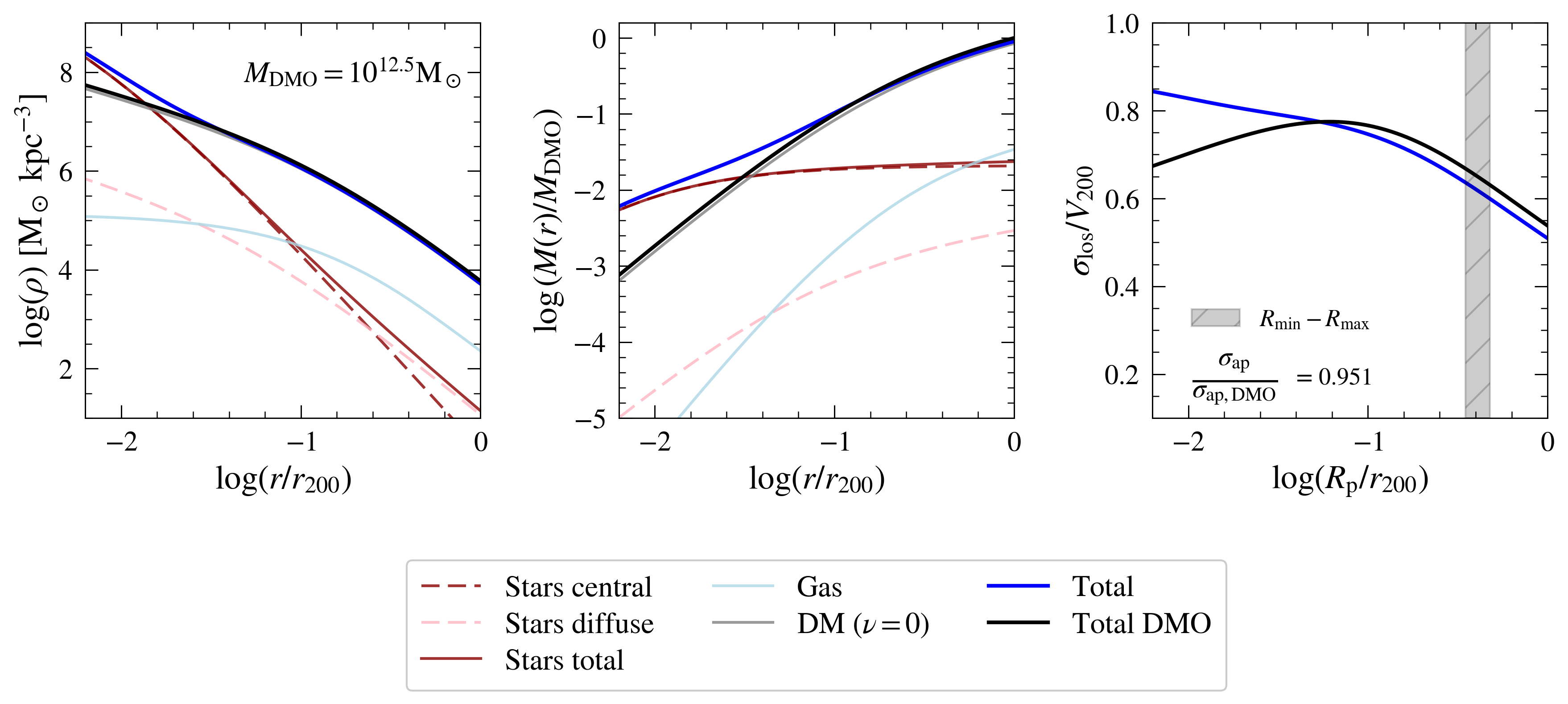}
    \includegraphics[width=\linewidth, trim = {0cm 0.2cm 0cm 0cm}, clip]{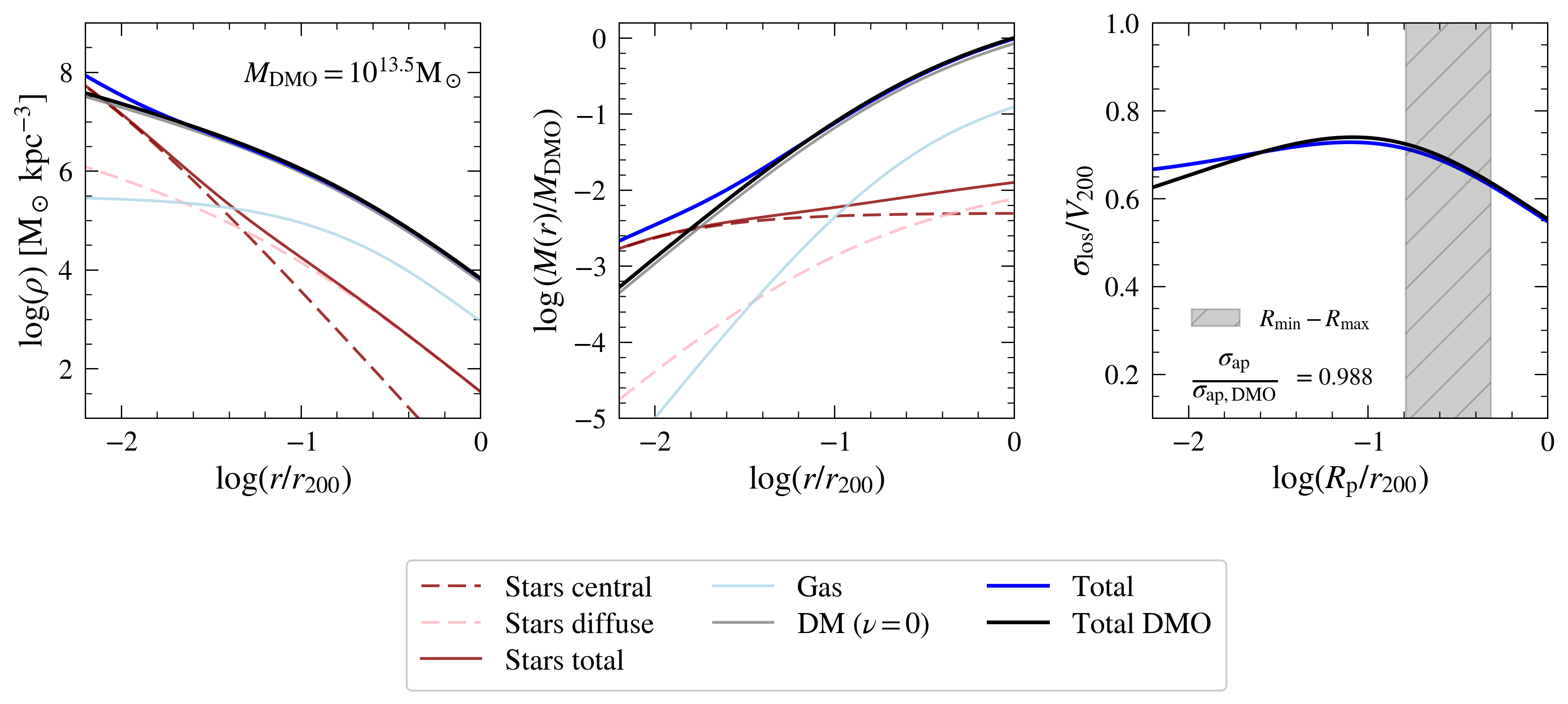}
    \caption{\textit{Left:}
    Analytical density profiles for two halos of mass $M_{\rm  DMO} = 10^{12.5}\mathrm{M}_{\odot}$ (top) and  $M_{\rm  DMO} = 10^{13.5}\mathrm{M}_{\odot}$ (bottom) for our fiducial model. We show the different components of the model; the total DMO profiles (black), dark matter in the baryonic model with no adiabatic response ($\nu=0$) (gray), the central galaxy (red dashed), diffuse stellar component (light pink dashed), total stellar component (red solid), gas (light blue) and total profile in the baryon model (blue).  
    \textit{Middle}: The corresponding enclosed mass profiles for the same two halos.
    \textit{Right}: The line-of-sight velocity dispersion for both the DMO model (black) and the baryonic model (blue), which are obtained from their corresponding density (equiv. mass) profiles. We highlight the region where satellites are selected in \Basilisc,  between the minimum radius ($R_{\mathrm{min}}$ = 55$\arcsec$ due to fibre collisions) and maximum radius ($R_{\mathrm{max}}$ = 0.375 $R_{\mathrm{vir}}$) as the gray shaded area. The integrated aperture velocity dispersion is computed over the radial range [$R_{\mathrm{min}}, R_{\mathrm{max}}$] for both the baryonic model ($\sigma_{\mathrm{ap}}$) and DMO model ($\sigma_{\mathrm{ap,DMO}}$). The ratio of the integrated velocity dispersion between the two models is  indicated in the lower left corner. }
    \label{fig:results:density_mass_vel_fiducial}
\end{figure*}

\subsection{Fiducial Model}
\label{sec:results:fiducialmodel}

In this section, we quantify the effects of baryons on the density, mass, and velocity dispersion profiles for our fiducial model. Table~\ref{tab:methodology:halomodel} (right‑most column) lists all adopted components. Where possible, the model is tied to empirical scaling relations, while the remaining parameters are calibrated to the EAGLE simulation. Specifically, the mass fractions of the ejected baryons and the diffuse stellar component follow the analytic fits indicated by the solid curves in Fig.~\ref{fig:methodology:eagle_icl_gas_fits}. We also consider two cases for the halo response (see Section~\ref{sec:methodology:adiabaticresponse}), namely, no response ($\nu=0$) and standard adiabatic response ($\nu=1$). For every figure in this work, we specify the value of $\nu$ used.

Regarding the specifics of the aperture over which we compute the satellite kinematics, we follow \citet{Mitra2024} who used \Basilisk to analyze data from the SDSS galaxy redshift survey. They selected central-satellite pairs that cover the redshift range $0.034 \leq z\leq 0.184$. Only satellites with a projected distance from the central given by $R_{\rm min} \leq R_\rmp \leq R_{\rm max}$ were used. Here $R_{\rm min}$ is taken to be equal to the fibre collision scale of the SDSS, which corresponds to $55 \arcsec$ \citep{Blanton2003}, while $R_{\rm max}$ scales with the luminosity of the central galaxy such that it roughly corresponds to a fixed fraction ($\sim 0.3$ to $0.4$) of the halo virial radius \citep[see also][]{vandenBosch2004, More2011, Lange2019b}. At $z=0.1$, the median redshift of the data used by \citet{Mitra2024}, $55\arcsec$ corresponds to $\sim 105\kpc$, which is the value we adopt here for $R_{\rm min}$, while throughout we set $R_{\rm max} = 0.375 \rvir$. 
\begin{figure*}
    \centering
    \includegraphics[width=\linewidth]{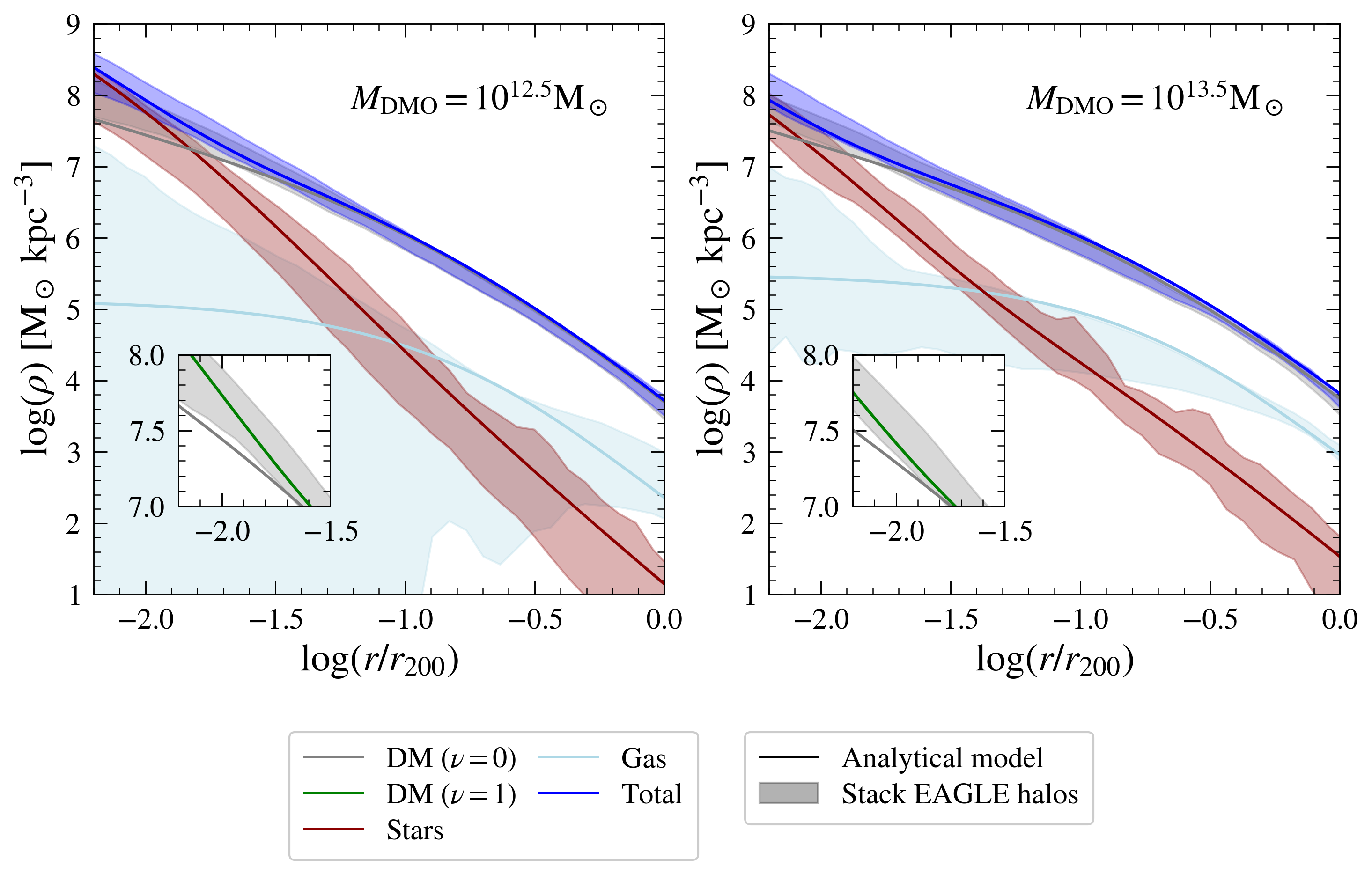}
    \caption{Comparison between the fiducial halo model and EAGLE halos of similar mass.  We show the density profiles for two halos of mass $M_{\rm  DMO} = 10^{12.5}\mathrm{M}_{\odot}$ (left) and  $M_{\rm  DMO} = 10^{13.5}\mathrm{M}_{\odot}$ (right). Solid lines represent the components of our halo model: dark matter (gray), total stellar mass (central + diffuse) (red), gas (light blue), and total density profile (blue). 
    To quantify the halo-to-halo variation in EAGLE halos, we compute the central 95\% spread in density in bins of $r/r_{200}$ for two stacks of EAGLE halos whose median DMO masses match the model: $10^{12.25}<M_{\rm  DMO}/\Msun<10^{12.85}$ (480 halos, left panel) and $10^{13.3}<M_{\rm  DMO}/\Msun<10^{13.9}$ (31 halos, right panel). This spread is shown as shaded regions in matching colors for each component. 
    In the main panels, the analytical model does not include halo response ($\nu=0$).  The insets in the lower-left corner zoom in on the inner dark matter density profiles: the gray curve shows the $\nu=0$ model (as in the main panel), while the green curve includes adiabatic response ($\nu=1$). } 
    \label{fig:results:stack_eagle}
\end{figure*}

The density profiles for the fiducial model are shown in Fig.~\ref{fig:results:density_mass_vel_fiducial} (left panels), along with their corresponding mass profiles (middle panels) and the line-of-sight velocity dispersion profiles (right panels) for a halo with $M_{\rm DMO}=10^{12.5}\Msun$ (top) and a halo with $M_{\rm DMO}=10^{13.5}\Msun$ (bottom).  In the inner region ($r\sim$0.01$r_{\rm 200}$) of the halo, the density (and mass) profiles are enhanced due to the presence of the central galaxy. As a result, the line-of-sight velocity dispersion is higher in this region than in the DMO model. Note that this effect is more pronounced for the halo with $M_{\rm DMO}=10^{12.5}\Msun$. This is a consequence of the fact that the SHMR peaks at halo masses of $M_{\rm DMO} \sim10^{12} \Msun$.

At larger radii, baryonic effects slightly reduce the density and enclosed mass profiles. This is a result of two effects: on the one hand, some of the mass that was originally in the outskirts has condensed towards the center to build the central galaxy; on the other hand, some mass is removed from the halo entirely, as quantified by $f_{\rm eject}(\MDMO)$. The grey, vertical bands in the right-hand panels indicate $R_{\rm min} \leq R_\rmp \leq R_{\rm max}$, the range of projected radii over which \citet{Mitra2024} probes the line-of-sight velocities of satellite galaxies. Note that, in this range of projected radii, the differences between the line-of-sight velocity profiles of the fiducial and DMO models are small and with a negligible dependence on $R_\rmp$. In particular, the fact that satellites with $R_\rmp < R_{\rm min} = 55''$ are excluded significantly reduces the sensitivity to baryonic effects, which typically manifest at smaller radii (see  Section~\ref{sec:fibre} for a more detailed discussion). 

For halos of $10^{12.5}\Msun$ ($10^{13.5}\Msun$), our fiducial model indicates that the line of sight velocity dispersion measured over an aperture with $R_{\rm min} \leq R_\rmp \leq R_{\rm max}$ is reduced by 5 (1) percent, relative to that in a DMO case. This reduction mainly reflects the reduction of the total halo mass due to the ejection of baryons, which is stronger in less massive halos. 

\subsection{Comparison with EAGLE}
\label{sec:results:comparisoneagle}

Fig.~\ref{fig:results:stack_eagle} compares the fiducial model to halos in EAGLE for two different masses: $M_{\rm DMO}=10^{12.5}\Msun$ (left panel) and $M_{\rm DMO}=10^{13.5}\Msun$ (right panel). Solid lines show the components of the analytical model, with different colors corresponding to different mass components, as indicated. The shaded regions in matching colors for each component indicate the halo-to-halo variation in EAGLE halos. 

%To quantify the halo-to-halo variation in EAGLE halos, we compute the central 95\% spread in density in bins of $r/r_{200}$ for two stacks of EAGLE halos whose median DMO masses match the model: $10^{12.25}<M_{\rm  DMO}/\Msun<10^{12.85}$ (480 halos, left panel) and $10^{13.3}<M_{\rm  DMO}/\Msun<10^{13.9}$ (31 halos, right panel). This spread is shown as shaded regions in matching colors for each component. 

Focusing on the individual halo components, we see that for $M_{\rm DMO}=10^{12.5}\Msun$, the analytical stellar mass profile of the fiducial model lies toward the high end of the stacked stellar mass profiles of EAGLE halos. This reflects the differences between the SHMR of \citet{Moster2013} on which the fiducial model is based, and that of the EAGLE simulations (see Fig.~\ref{fig:methodology:eagle_icl_gas_fits}). Note, though, that the analytical model falls well within the halo-to-halo scatter of the simulation results. Although the gas profiles in EAGLE exhibit significant scatter, particularly in the central regions, the overall trend is broadly reproduced by our fiducial analytical model. For more massive halos, the scatter decreases, and the agreement improves, especially at larger radii where the gas distribution is well-approximated by an NFW profile. The dark matter profiles in EAGLE halos exhibit central contraction, an effect not included in the main panels. However, allowing for contraction with $\nu=1$ (see Section~\ref{sec:methodology:adiabaticresponse}) yields very good agreement with the EAGLE profiles, as shown by the green curves shown in the insets in the bottom-left.

Finally, the total density profiles show good agreement between the EAGLE simulations and our fiducial analytical model. This is noteworthy because, although the simulation data was used to calibrate $f_{\rm \ast, diffuse}(\MDMO)$ and $f_{\rm eject}(\MDMO)$, the model was not tuned to reproduce any of the density profiles in EAGLE. Since the kinematics of the satellite galaxies are ultimately controlled by the total enclosed mass profile, we can thus be confident that our fiducial analytical model is meaningful.

\subsection{Variations from the fiducial model}
\label{sec:results:variations_from_fiducial}

In this section, we deviate from the fiducial model by varying free parameters in the baryonic model to probe the full range of baryonic effects on satellite kinematics. We find that, to good approximation, the impact of baryons is dominated by only two parameters; the halo response, $\nu$, and the ejection strength, $f_{\rm eject}$, This is a result of the fact that the fiducial model excludes the innermost 55$\arcsec$ from the analysis, effectively removing sensitivity to baryonic effects that influence the central regions.  As a result, the radial range probed in our fiducial model, which is in agreement with the SDSS-based analysis of \Basilisc, is governed almost entirely by $f_{\rm eject}$ and $\nu$, whose effects on the velocity dispersion we discuss below. Other variations, such as changes in the central galaxy or the radial profile of the satellites, are addressed in Appendix~\ref{sec:appendix:negligible_parameters} and are shown to have a negligible impact. This insensitivity would not persist if a significantly smaller minimum aperture radius were adopted, as discussed in Section~\ref{sec:fibre}.

\begin{figure*}
    \centering
    \includegraphics[width=\linewidth]{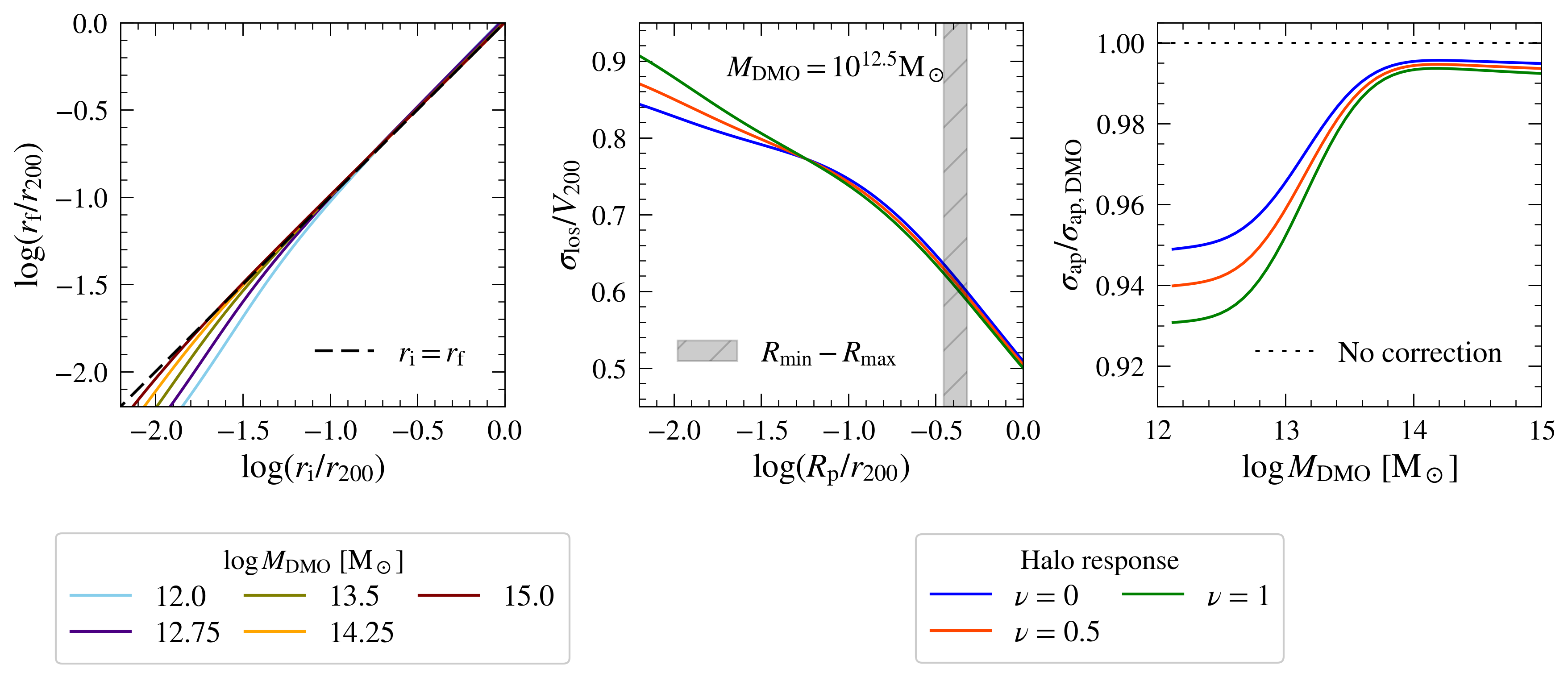}
    \caption{Illustrations of the halo response. \textit{Left}: Solving for the response of the dark matter due to the presence of the baryons for the fiducial model with $\nu=1$ for different halo masses. The final radius $r_{\rm f}$ is plotted against initial radius $r_{\rm i}$. The dashed line shows $r_{\rm f}= r_{\rm i}$, indicating when there would be no halo response. For small $r_{\rm i}$ we see that $r_{\rm f}<r_{\rm i}$, indicating that dark matter that was initially at larger radii migrated inwards (contraction). On the other hand, at at large $r_{\rm i}$, we see that $r_{\rm f}\geq r_{\rm i}$, indicating a very moderate expansion. Both effects become weaker with increasing halo mass. \textit{Middle}: The line-of-sight velocity dispersion profile for a halo of mass $M_{\rm DMO}\sim10^{12.5}\Msun$, for $\nu=0$ (blue), $\nu=0.5$ (red) and $\nu=1$ (green). Under $\nu>0$, we again see that the dark matter halo responds by contracting in the center and expanding in the outskirts, though only the latter affects satellite kinematics data (gray shaded region).  
    \textit{Right}: Impact of the halo response on the aperture velocity dispersion ratio, $\sigma_{\rm ap}/\sigma_{\rm ap, DMO}$, as a function of halo mass. The blue curve indicates no halo response $\nu=0$, equivalent to $r_{\rm i}=r_{\rm f}$. When the dark matter responds ($\nu>0$), the observed effect is a further decrease of $\sigma_{\rm ap}/\sigma_{\rm ap, DMO}$.}
    \label{fig:results:adiabaticresponse}
\end{figure*}

\subsubsection{Varying halo response strength}
\label{sec:results:adiabaticresponse}

The left panel of Fig.~\ref{fig:results:adiabaticresponse} shows solutions to equation~(\ref{eq:methdology:adiabatic_invariant}), which determines how the dark matter profile adjusts to preserve its adiabatic invariants. The plot shows the final radius, $r_\rmf$, of a spherical shell of dark matter, as a function of the shell's initial radius, $r_\rmi$, in response to the process of galaxy formation. Here we have adopted the standard adiabatic response formalism with $\nu=1$, and our fiducial model for the various baryonic components. Note that the relation $r_\rmf(r_\rmi)$ has a clear dependence on halo mass, as indicated by the different colors. For small $r_{\rm i}$ (inner region), the final radius is smaller than the initial radius, indicating that the central halo responds by contracting. This is simply a consequence of the accumulation of baryonic matter at the halo center associated with the formation of the central galaxy. For large $r_{\rm i}$, the opposite effect occurs and the final radius is slightly larger than the initial radius, indicating that the outer halo responds by expanding. This is a consequence of the ejection of baryons, which makes the remaining matter less strongly bound. Note that both of these effects, central contraction and expansion in the outskirts, become weaker with increasing halo mass. This is because both the ratio $M_{\mathrm{\ast, cen}}/M_{\rm DMO}$ and the ejected mass fraction decrease with increasing $M_{\rm DMO}$. 

The middle panel of Fig.~\ref{fig:results:adiabaticresponse} plots the line-of-sight velocity dispersion profile for a halo of mass $M_{\rm DMO}\sim10^{12.5}\Msun$, for $\nu=0$ (blue), $\nu=0.5$ (red) and $\nu=1$ (green). The case of $\nu=0$ is identical to the blue curve in the top right panel of Fig.~\ref{fig:results:density_mass_vel_fiducial}. In the most extreme case, $\nu=1$, we have maximal contraction in the inner region, plus maximal expansion in the outer regions. The shaded gray area again represents the radial range of satellite kinematics measurements, highlighting that although adiabatic contraction can have a strong impact in the central regions, the dominant effect impacting the data is dark matter expansion. As a result, the primary impact of halo response is a reduction of the aperture velocity dispersion ratio, $\sigma_{\rm ap}/\sigma_{\rm ap, DMO}$. The right panel of Fig.~\ref{fig:results:adiabaticresponse} shows $\sigma_{\rm ap}/\sigma_{\rm ap, DMO}$ as a function of halo mass. The blue curve corresponds to no response, $\nu = 0$, equivalent to $r_{\rm i} = r_{\rm f}$.  We also show the curves for $\nu =0.5$ (red) and $\nu=1$ (green).  The strongest effect is observed for $M_{\rm DMO}\sim10^{12.5}\Msun$. For these halo masses, $f_{\rm eject}$ is large, and the halo response has a significant impact, amplifying the effects of mass loss by further reducing $\sigma_{\rm ap}/\sigma_{\rm ap, DMO}$ by an additional $\sim$2\%. 
\begin{figure*}
    \centering
    \includegraphics[width=\linewidth]{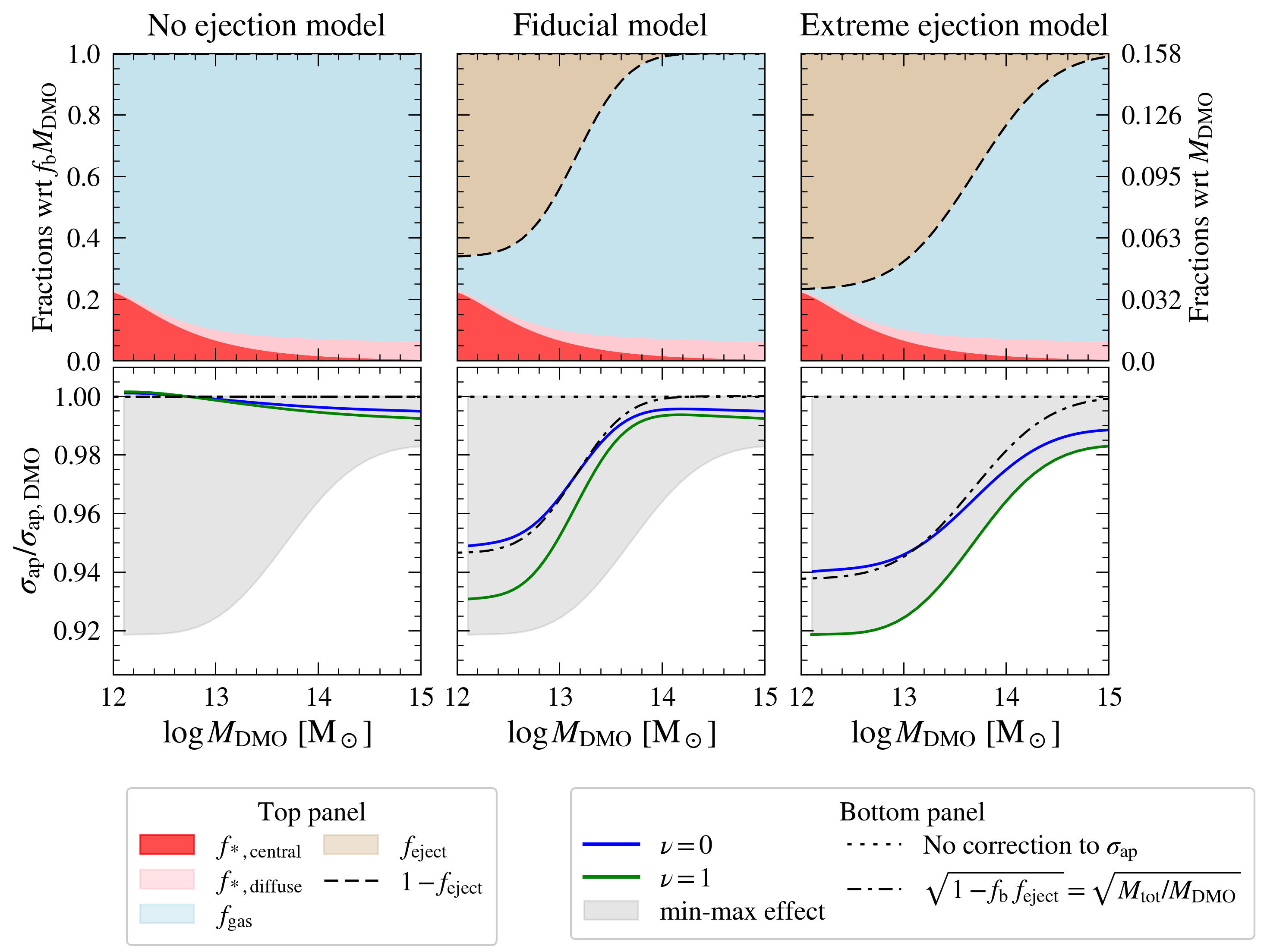}
    \caption{  
    Comparison of the effects of the different ejection models. \textit{Top panel}: The cumulative mass fractions of each baryonic component as a function of halo mass for the three scenarios considered (see text): the 'no-ejection model' (left), the fiducial model (middle), and the 'extreme-ejection model' (right). In each scenario, we keep the mass fractions of the central galaxy (red) and diffuse stellar component (pink) fixed, and we only vary the parameter $f_{\rm eject}$ (brown), which leads to different total gas fractions (blue). The total bound baryonic fraction, shown as the black dashed line, is simply obtained by summing all bound baryonic fractions, which is by definition equal to 1 minus the ejected fraction. 
   \textit{Bottom panel}: the impact of baryonic effects on the ratio of integrated velocity dispersion $\sigma_{\rm ap}/\sigma_{\rm ap, DMO}$ as a function of halo mass. Without baryonic corrections, the ratio remains at 1 (dotted black line). The blue curves, assuming no dark matter halo response, follow roughly $\sigma_{\rm ap}/\sigma_{\rm ap, DMO} \approx $ $\sqrt{M_{\rm tot}/M_{\rm DMO}} = \sqrt{1-f_{\rm b} f_{\rm eject}}$ (dashed black line). The green curves include halo response when preserving adiabatic invariants ($\nu=1$), further reducing velocity dispersion ratio. The gray shaded region represents the range defined by the extreme models, spanning zero to maximal baryonic effect.}
    \label{fig:results:finalplot}
\end{figure*}

\subsubsection{Varying ejection strength}
\label{sec:results:extreme_models}

In this section, we examine the impact of changes of the parameters of the function $f_{\rm eject}(\MDMO)$ given in equation~(\ref{eq:methodology:feject_function}). Since mass conservation requires $f_{\rm gas} + f_{\rm *,cen} + f_{\rm *, diffuse} + f_{\rm eject} = 1$, such changes also imply adjustment to one or more other mass fractions. To fully understand the consequences of these trade-offs, we test multiple configurations and define two limiting models that represent extreme, yet self-consistent, baryonic configurations.  

In the first extreme scenario, we keep $f_{\rm *,cen}$ and $f_{\rm *, diffuse}$ at their fiducial values, but set $f_{\rm eject}=0$. This means that we are not ejecting {\it any} baryons (e.g. no stellar or AGN outflows); for halos at all mass scales, the gas now complements the stars such that the total bound baryon fraction adds up to the universal baryon fraction. Since the gas contributes to the mass in the outskirts of the halo, where it has a density profile that is similar to that of the dark matter, this model leads to minimal baryonic effects on satellite kinematics. We refer to this configuration as the `no-ejection' model.

In the second extreme model, we again keep the fiducial fractions for the central galaxy and diffuse stellar component, but we change the gas ejection parameters that describe the mass dependence of $f_{\rm eject}$ to $\alpha = 1.2$, $\log M_{\mathrm{char, eject}}=13.7$ and $f_{\mathrm{eject, max}} = 0.8$. These values give $f_{\rm eject} \sim 0.8$ at $10^{12} M_{\odot}$, which is the extreme for which all gas is blown out ($f_{\rm gas}=0$) in halos of this mass. In addition, the changes in $\alpha$ and $M_{\mathrm{char, eject}}$ lead to a slower decrease in the ejected mass fraction with halo mass, which increases the ejected mass fractions in all halos. However, we ensure that $f_{\rm eject}=0$ for halos with $M_{\rm DMO} \geq 10^{15} M_{\odot}$. We refer to this as the `extreme-ejection' model. We also set the transition radius for the gas profile at $x_{\rm tr}=c_{200}$ (see Appendix~\ref{sec:appendix:negligible_parameters}).This configuration induces the strongest baryonic effects on satellite kinematics. 

The top panels of Fig.~\ref{fig:results:finalplot} show the cumulative mass fractions of the baryonic components as a function of halo mass for the three models considered here: the `no-ejection' model (left), the fiducial model (middle), and the `extreme-ejection' model (right). In all three cases, the mass fractions of the central galaxy and the diffuse stellar component are identical, the only variation being the ejected mass fraction, $f_{\rm eject}(M_{\rm DMO})$, which leads to different gas fractions. The total bound baryonic fraction is the sum of the mass fractions of the central galaxy, diffuse stars and gas, which is by definition equal to $1-f_{\rm eject}$. In the no-ejection model, the bound baryonic fraction matches the universal baryon fraction across all halo mass scales. In the fiducial model, in the absence of halo response to baryons ($\nu=0$), the bound baryonic fraction is much lower ($\sim$ 0.3) at mass scales of $10^{12}-10^{13}\Msun$ but rises to match the universal baryon fraction at $M_{\rm DMO} \sim 10^{14}\Msun$. Finally, the extreme-ejection model has a baryon mass deficit at all scales, reaching the universal baryon fraction only at a halo mass scale of $10^{15}\Msun$. 
\begin{figure*}
    \centering
    \includegraphics[width=0.95\linewidth]{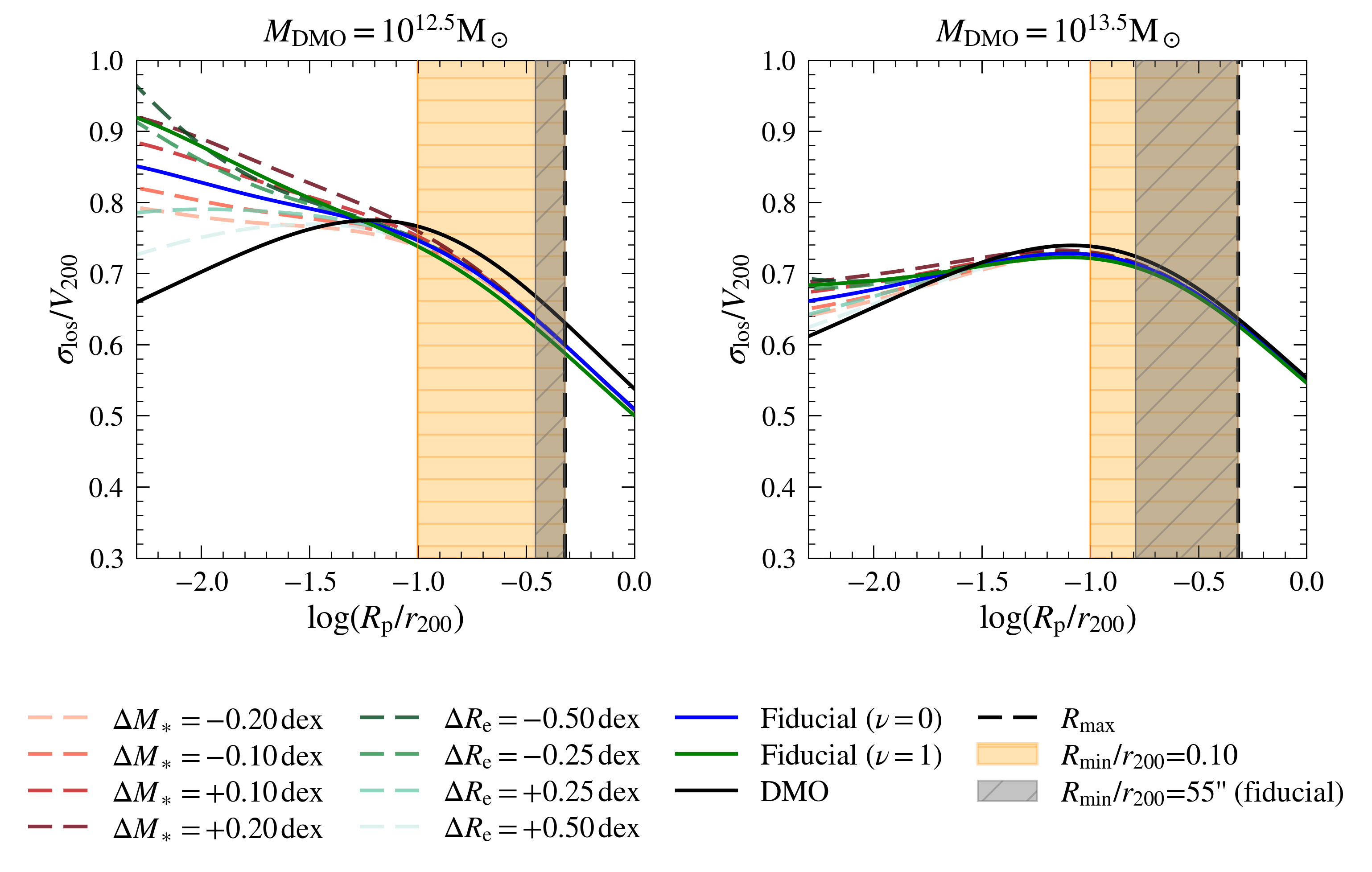}
    \caption{Line-of-sight velocity dispersion profiles for halos of $10^{12.5} \Msun$ (left) and $10^{13.5} \Msun$ (right), showing the impact of varying the scatter in the stellar mass–halo mass relation ($\Delta M_{\rm *}$) and the size–stellar mass relation ($\Delta R_{\rm e}$). The solid blue curve shows the fiducial model without halo response ($\nu=0$), while the solid green curve corresponds to the same fiducial baryonic configuration but including adiabatic halo response ($\nu=1$).
    For all variations in $\Delta M_{\rm *}$ and $\Delta R_{\rm e}$ the halo response parameter $\nu=0$.
    The shaded regions indicate the radial ranges used to compute the aperture velocity dispersion, corresponding to two values of $R_{\rm min}$: the fiducial 55$\arcsec$ fibre collision scale (grey) and $0.1 r_{200}$ (yellow). While stellar properties significantly affect the velocity dispersion at small radii, the fiducial aperture, set by the SDSS fibre collision scale, fortunately excludes these central regions, simplifying the baryonic corrections.}
    \label{fig:results:Rmin}
\end{figure*}

The bottom panels of Fig.~\ref{fig:results:finalplot} show the corresponding $\sigma_{\rm ap}/\sigma_{\rm ap, DMO}$ ratios as a function of halo mass. In the absence of baryonic effects,  $\sigma_{\rm ap}/\sigma_{\rm ap, DMO}=1$, indicated with the black dotted line. The blue curves show the impact of baryonic effects under the assumption of no adiabatic response of dark matter ($\nu=0$). In the no-ejection model, the total mass within the halo virial radius is conserved, such that $M_{\rm tot}=M_{\rm DMO}$. Small deviations ($\lta 0.5\%$) of $\sigma_{\rm ap}/\sigma_{\rm ap, DMO}$ from unity are apparent, and due to the fact that the baryonic matter is distributed differently from that of the dark matter. Nevertheless, it is clear that if all halos were to retain their baryons, the baryonic effects on satellite kinematics are virtually negligible. For the fiducial model, $\sigma_{\rm ap}/\sigma_{\rm ap, DMO} \simeq 0.95$ at the low mass end ($M_{\rm DMO}\sim10^{12}\Msun$), but gradually increases to $\sigma_{\rm ap}/\sigma_{\rm ap, DMO} \simeq 0.995$ for $M_{\rm DMO} \gta 10^{14}\Msun$. In the extreme-ejection model, the ratio further decreases to approximately $\sigma_{\rm ap}/\sigma_{\rm ap, DMO}=0.94$ at $M_{\rm DMO} \sim 10^{12}\Msun$, increasing to $\sigma_{\rm ap}/\sigma_{\rm ap, DMO}=0.99$ at $M_{\rm DMO} \sim 10^{15}\Msun$. An additional 0.5\% reduction at the high-mass end results from setting $x_{\rm tr} = c_{200}$, which shifts the transition of the gas profile from the polytropic to the NFW profile to a radius twice as large as in the fiducial model (see Appendix~\ref{sec:appendix:negligible_parameters}).

The green curves show the halo mass dependence of $\sigma_{\rm ap}/\sigma_{\rm ap, DMO}$ in the cases where we account for the response of dark matter using the standard adiabatic invariance formalism with $\nu=1$. As discussed in the previous section, mass loss causes the dark matter to expand in its outskirts, leading to a further reduction in $\sigma_{\rm ap}$, which is most pronounced at $M_{\rm DMO} \sim 10^{12}-10^{13}\Msun$.

To first order, the velocity dispersion of satellite galaxies squared simply scales with the dynamical mass. Hence, we expect $\sigma_{\rm ap}/\sigma_{\rm ap, DMO} \approx \sqrt{M_{\rm tot}/M_{\rm DMO}}$. And since $M_{\rm tot}/M_{\rm DMO} = 1 - f_\rmb f_{\rm eject}$, we expect that $\sigma_{\rm ap}/\sigma_{\rm ap, DMO}$ is mainly determined by $f_{\rm eject}(M_{\rm DMO})$. The black dot-dashed curves in the bottom panels of Fig.~\ref{fig:results:finalplot} show the predicted $\sigma_{\rm ap}/\sigma_{\rm ap, DMO} = \sqrt{1 - f_\rmb f_{\rm eject}}$. Note that these predictions are in reasonable agreement with the actual ratio $\sigma_{\rm ap}/\sigma_{\rm ap, DMO}$, but only in the case of no halo response ($\nu=0$, indicated by the blue curves). With halo response, these simple predictions overestimate $\sigma_{\rm ap}/\sigma_{\rm ap, DMO}$, by between $0.5$ and 2 percent.

Finally, using the two extreme models introduced here, we can conservatively outline the uncertainties in $\sigma_{\rm ap}/\sigma_{\rm ap, DMO}$ resulting from our incomplete understanding of galaxy formation. This is indicated by the gray-shaded regions in the lower panels of Fig.~\ref{fig:results:finalplot}. It basically spans the range of results from zero baryonic effects (roughly what is expected if no baryons are ever ejected), to the results of the extreme-ejection model combined with maximal halo response ($\nu = 1$). It encompasses the fiducial curve and, as demonstrated in Appendix~\ref{sec:appendix:negligible_parameters}, also captures the effects of variations in the central galaxy, diffuse stellar component, satellite profile, dark matter concentration, and redshift. 

%\newpage
\subsection{The (fortuitous) impact of fibre collisions}
\label{sec:fibre}

The fiducial model used in this work -- designed to match the SDSS satellite selection in \Basilisk -- uses an aperture that excludes the inner 55$\arcsec$ due to fibre collisions in the spectroscopic data. At the median redshift of the SDSS sample ($z=0.1$), this corresponds to a projected physical scale of approximately 100 kpc. This exclusion turns out to be remarkably fortunate, since several baryonic effects predominantly impact the kinematics in these inner regions.  By excluding these regions from the analysis, the modelling is naturally shielded from the complexities associated with such effects, as demonstrated in Appendix~\ref{sec:appendix:negligible_parameters}. 

Figure~\ref{fig:results:Rmin} further illustrates this by showing the line-of-sight velocity dispersion profiles for halos of $10^{12.5} \Msun$ (left panel) and $10^{13.5} \Msun$ (right panel), while varying the mass of the central galaxy at fixed halo mass by up to 0.2 dex, reflecting the typical scatter in the SHMR, and the size of the central galaxy at fixed stellar mass by up to 0.5 dex, roughly the scatter in the galaxy size-mass relation. As expected, central galaxies that are more massive or more compact increase the line-of-sight velocity dispersion, but only for $R_\rmp \lta 0.1 r_{200}$. If satellite kinematics data were to probe these inner regions, the baryonic corrections would become extremely sensitive to the detailed structure (stellar mass and size) of the central galaxy and the response that its formation induces in the central halo. This would significantly complicate constraining the galaxy-halo connection using satellite kinematics. Hence, the fact that the SDSS data are affected by fibre collisions, which prompted \citet{Mitra2024} to ignore any data inside of 55$\arcsec$ from the centrals, is a rather fortuitous imperfection of the data. In fact, our study suggests that any future analysis of satellite kinematics data is better off excluding potential data of satellite galaxies with a projected separation $R_\rmp \lta 0.1 r_{200}$, as the increased complexity and uncertainties associated with baryonic corrections are likely to outweigh the gain in statistical power. 

\begin{figure*}
    \centering
    \includegraphics[width=0.98\linewidth]{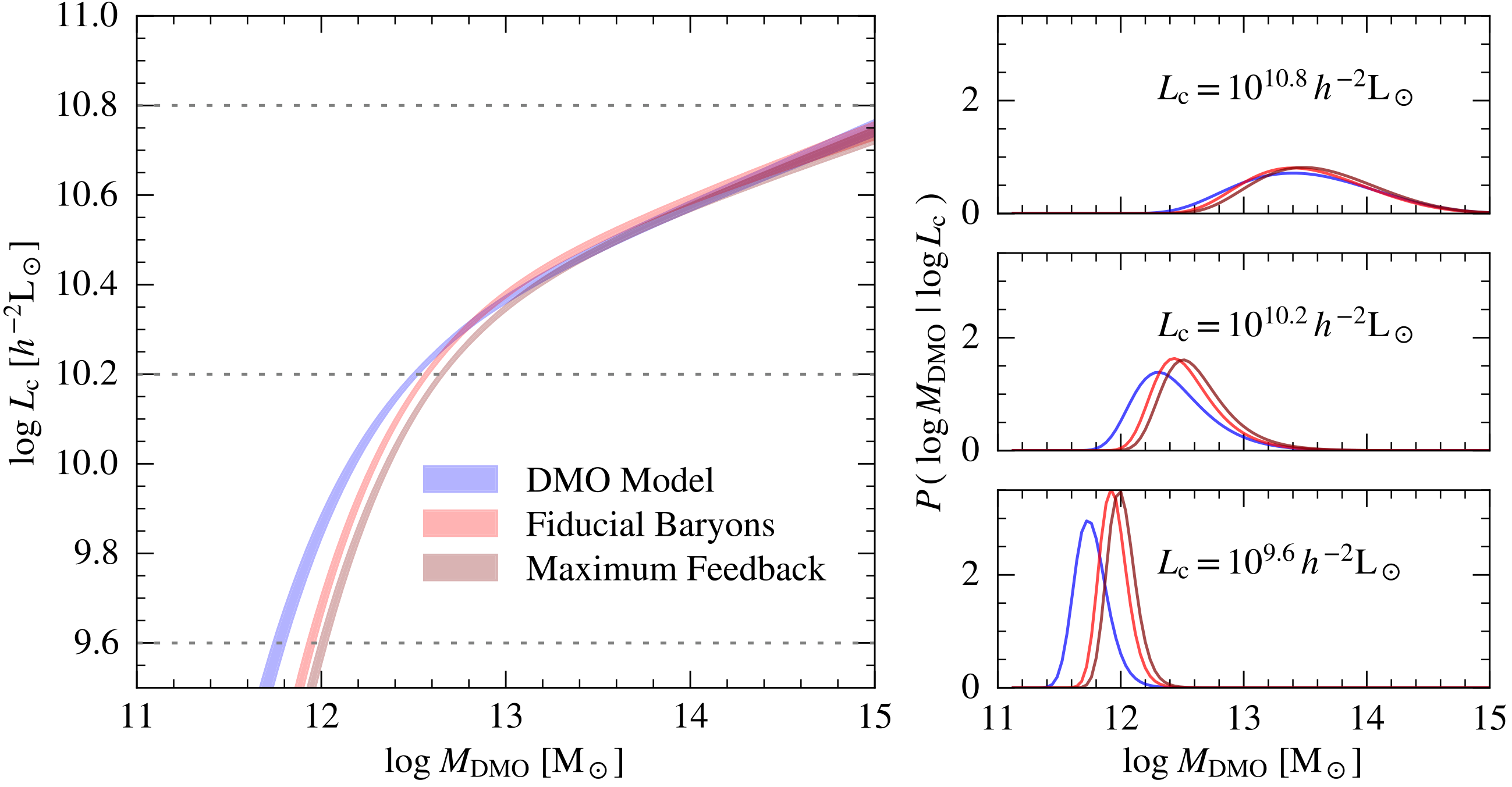}
    \caption{\textit{Left}: Luminosity of central galaxies as a function of halo mass, as inferred from \Basilisk \citep{Mitra2024} with no baryonic correction (DMO model, blue), with the fiducial baryonic model plus adiabatic response of the dark matter (pink), and with the maximum feedback model along with adiabatic response (brown). Each shaded band represents the 68\% confidence interval of the median central luminosity-halo mass relation. \textit{Right}: Halo mass posterior at a given central galaxy luminosity for three values; $L_{\rm c}=10^{10.8} h^{-2} \Lsun$ (top), $L_{\rm c}=10^{10.2} h^{-2} \Lsun$ (middle) and $L_{\rm c}=10^{9.6} h^{-2} \Lsun$ (bottom).}
    \label{fig:bias_dynamical_mass}
\end{figure*}

\subsection{Bias on dynamical mass inference due to baryonic effects}

We have explored in detail how baryonic effects influence the kinematics of satellite galaxies. In this section, we examine how these effects propagate into the inference of halo masses and, consequently, the galaxy–halo connection. To include baryonic effects in \Basilisk, we rescale the line-of-sight velocity dispersion of satellites, predicted assuming a halo composed entirely of dark matter following an NFW profile, by the halo-mass–dependent ratio $\sigma_{\rm ap}/\sigma_{\rm ap,DMO}$ derived in this work, considering both the fiducial model and the extreme ejection model with adiabatic response (green curves in Fig.~\ref{fig:results:finalplot}). Unlike \citet{Mitra2024}, we use \Basilisk to fit only the kinematics and abundance of satellite galaxies, excluding the luminosity function (LF) as a separate constraint. This allows us to isolate and illustrate the specific impact of baryonic effects on dynamical mass inference. 

In Fig.~\ref{fig:bias_dynamical_mass}, we present the inferred galaxy–halo connection under three scenarios: (1) no baryonic correction (DMO) (blue), similar to \citet{Mitra2024}, but without using the LF as additional constraints; (2) using the fiducial baryonic correction model (pink), and (3) using the extreme ejection model (brown). The left panel presents the median relation between central galaxy luminosity and DMO halo mass. Each shaded region indicates the corresponding 68\% confidence interval in the inferred $L_{\rm c}-M_{\rm DMO}$ relation. The right panels show the best-fit halo mass distributions at fixed central galaxy luminosity for three cases: $L_{\rm c}=10^{10.8}\,h^{-2}\Lsun$ (top), $L_{\rm c}=10^{10.2}\,h^{-2}\Lsun$ (middle), and $L_{\rm c}=10^{9.6}\,h^{-2}\Lsun$ (bottom).
We note that for the highest luminosity bin (top right panel), the posterior peaks at a lower mass than indicated by the median relation in the left-hand panel (where $L_{\rm c} \approx 10^{10.8}\,h^{-2}\Lsun$ corresponds to $M_{\rm DMO} \sim 10^{15}\Msun$). This is not due to a prior on the halo mass function, but simply reflects the exponential suppression of the halo mass function at the high-mass end.

The differences between the DMO model and the two baryonic models are most pronounced at the low-mass end. This is expected, because in the low-mass end $f_{\rm eject}$ is the largest in the baryonic models, leading to the strongest suppression of the enclosed mass profile in individual halos. Thus, for a given $M_{\rm DMO}$, halos in the fiducial feedback scenario exhibit lower velocity dispersions compared to the DMO case. Conversely, for a given observed velocity dispersion, halos in the baryonic scenario require a higher $M_{\rm DMO}$, compared to the DMO case, to match the same satellite kinematics after feedback-driven suppression. As a result, including the baryonic corrections systematically shifts the inferred $M_{\rm DMO}$ to higher values for central galaxies at fixed luminosity. The shift is larger in the maximum feedback scenario, as expected, but the difference relative to the fiducial case is minimal.
%with both scenarios yielding shifts of up to $\sim$0.2~dex at the low-luminosity end.
%\new{This effect is more enhanced for the maximum feedback scenario, as expected, although the difference is not very significant}. 

Overall, baryonic effects seem to have a fairly negligible impact on the galaxy-halo connection inferred from satellite kinematics for halos above $10^{13} \Msun$. However, for halos with masses around $10^{12} \Msun$, comparable to that of the Milky Way, not accounting for baryonic effects causes the inferred median luminosity of centrals to be overestimated by as much as $0.3$ dex. As demonstrated in a subsequent paper \citep{Mitra2025arXiv251214889M_Basilisk_no_tension}, such systematic effects can propagate to significant systematic errors in cosmological inference.

\section{Summary and Conclusion}
\label{sec:discussion}

The kinematics of satellite galaxies are a powerful probe of the dynamical masses of the halos in which the satellites orbit. In particular, the Bayesian hierarchical inference method \Basilisc, recently developed by \citet{vandenBosch2019} and~\citet{Mitra2024}, has been demonstrated effective in using satellite kinematics data, extracted from a large galaxy redshift survey such as the SDSS, to put constraints on the galaxy-halo connection. These constraints are competitive with and complementary to other methods, such as galaxy clustering and galaxy-galaxy lensing,  while having the additional advantage of being insensitive to halo assembly bias. However, virtually all studies based on satellite kinematics to date, including those with \Basilisc, have relied on dark matter-only (DMO) halo models, neglecting the role of baryons, which can systematically bias the inferences.

In this study, we therefore developed an analytical halo model to quantify the impact of baryonic effects on satellite kinematics. We introduced two models for the halo's density profile: a dark matter-only (DMO) model and a model with baryons, which serves as its counterpart, incorporating stars, gas, and the adiabatic response of dark matter. We then compute the line-of-sight velocity dispersion, integrated over an aperture, for the model with baryons ($\sigma_{\rm ap}$), and compare it to the corresponding value obtained from the DMO model ($\sigma_{\rm ap, DMO}$). The ratio $\sigma_{\rm ap}/\sigma_{\rm ap, DMO}$ as a function of DMO halo mass defines a baryonic correction function that is easily incorporated in analyses of satellite kinematics data. Although we have applied the model specifically within the context of \Basilisc, its broader applicability makes it a useful framework for other analyses that rely on the one-halo term, including clustering and galaxy-galaxy lensing. A major strength of the analytical model is its flexibility, which enables a systematic identification of the baryonic components most responsible for the observed effects.

The most important parameter that drives the effect of baryons on satellite kinematics is $f_{\rm eject}$,  which controls the amount of baryons ejected from the halo relative to the maximum possible baryonic content. Setting $f_{\rm eject} = 0$, the gas and stellar fractions together match the universal baryon fraction, so the total bound halo mass in the baryonic model equals that of the DMO model. But when $f_{\rm eject} > 0$, the total halo mass in the baryon model will be reduced compared to the DMO model. Given that $f_{\rm eject}$ thus directly sets the total bound mass of the halo, its significance is entirely expected. The line-of-sight velocity dispersion, $\sigma$, scales as the square root of the enclosed mass, meaning any change in halo mass propagates directly to $\sigma$. Thus, the first-order correction for baryonic effects on satellite kinematics is to account for the change in bound halo mass. 

Being the dominant parameter, $f_{\rm eject}$ is also the most uncertain, with essentially no direct observational constraints on its functional form. In this work, we adopt $f_{\rm eject}$ from the median behavior of halos in the EAGLE simulation for our fiducial model, where low-mass halos experience significant baryon loss while high-mass halos retain their baryons. Although this trend of an increasing baryon fraction with halo mass is broadly consistent with limited observational evidence, the exact shape of $f_{\rm eject}$ in our fiducial model is ultimately determined by the specific feedback implementation in EAGLE. To capture the uncertainty in $f_{\rm eject}$, we explored two extreme scenarios. In the first scenario, we assume no ejection at any halo mass, such that the baryon content matches the universal baryon fraction across all scales. In this case, baryonic effects on satellite kinematics are suppressed to below the 1\% level. In the second, we assume maximal ejection at the low-mass end, with the baryon fraction gradually building up to the universal value only by $M_{\rm DMO}=10^{15}\Msun$. This results in the strongest baryonic effects, with a reduction in the velocity dispersion of about 6\% at the $M_{\rm DMO}=10^{12}\Msun$ scales. Although a detailed comparison with other simulations lies beyond the scope of this work, it would be informative to examine how $f_{\rm eject}$ behaves in different feedback models. In addition, future observational efforts targeting baryon fractions and the distribution of baryons, particularly in the CGM, as a function of halo mass will be crucial for constraining the exact form of $f_{\rm eject}$. 

The second most important effect is the adiabatic response of the dark matter. Although baryons can induce contraction of the dark matter at small radii, satellite kinematics predominantly probes the outer regions of halos. In these outskirts, a reduction in halo mass leads the dark matter to expand in order to conserve its adiabatic invariants. This expansion further suppresses the velocity dispersion ratio, $\sigma_{\rm ap}/\sigma_{\rm ap, DMO}$, with the effect being strongest around $M_{\rm DMO} \sim 10^{12.5}\Msun$. Consequently, the response of the dark matter halo acts as an amplifier of the primary baryonic effect: the more baryons are expelled, the greater the dark matter expansion, and the stronger the resulting reduction in the observed velocity dispersion.

Variations in any other stellar parameters such as the scatter in the stellar–halo mass relation, the galaxy size–stellar mass relation, and the properties of the diffuse stellar component have a negligible impact. This insensitivity stems from a key observational limitation: the SDSS fibre collision scale ($R_{\rm min} = 55\arcsec$) excludes the central regions of halos from the analysis. As a result, the measurements are primarily sensitive to the outer halo, where the dominant baryonic effects are governed by the reduced total mass set by $f_{\rm eject}$ and the halo response $\nu$. To ensure simplistic baryonic corrections, we recommend applying a comparable radial cut even when datasets are not subject to fibre collision limitations.

Finally, we implement the baryonic correction function developed here into the \Basilisk framework, following \citet{Mitra2024}, and demonstrate that accounting for baryonic effects leads to a systematic shift in inferred halo masses. Specifically, for low-mass halos where baryon ejection is most significant, the suppression in velocity dispersion results in underestimates of the inferred halo mass if uncorrected. Applying the correction shifts central galaxies to halos that are typically 0.1-0.2 dex more massive at fixed luminosity. 

In a companion paper \citep{Mitra2025arXiv251214889M_Basilisk_no_tension}, we demonstrate that such percent level changes in the inferred halo masses can significantly shift the cosmological constraints. This underscores the need to account for baryonic physics when inferring the galaxy–halo connection and deriving cosmological parameters. Moreover, as discussed in \citet{Mitra2025arXiv251214889M_Basilisk_no_tension}, one can invert this approach by comparing constraints from satellite kinematics corrected for baryonic effects with those from {\it Planck}.  This comparison opens a potential window for using satellite kinematics data to constrain $f_{\rm eject}$ as a function of halo virial mass, and thus the efficiency of feedback processes associated with galaxy formation.

\newpage
\section*{Acknowledgments}

We are grateful to Joop Schaye and Matthieu Schaller for their help with the EAGLE simulations. FvdB is supported by the National Science Foundation (NSF) through grants AST-2307280 and AST-2407063. KM's work at Argonne was supported under the DOE contract DE-AC02-06CH11357. This work used the following python packages: \texttt{Matplotlib} \citep{Matplotlib_Hunter2007}, \texttt{SciPy} \citep{SciPy_Virtanen2020}, \texttt{NumPy} \citep{numpy_vanderWalt2011}, \texttt{halotools} \citep{Hearin2017_halotools}, \texttt{AstroPy} \citep{Astropy2022} and \texttt{colossus} \citep{Diemer2018_colossus}.  
Finally, we are very grateful to the referee, Gary Mamon, for an extremely careful and detailed report that helped strengthen this work.

%%%%%%%%%%%%%%%%%
% Bibliography
%%%%%%%%%%%%%%%%%

\bibliographystyle{mnras}
\bibliography{paper}

%%%%%%%%%%%%%%%%%%%%%%%%%%%%%%%%%%%%%%%%%%%%%%%%%%%%%%%%%%%%%%

\appendix

\numberwithin{figure}{section}
\numberwithin{table}{section}
\numberwithin{equation}{section}

\section{Variations of parameters with minimal effects}
\label{sec:appendix:negligible_parameters}

Here, we explore variations in several additional free parameters of our model. Each parameter is varied individually, while the others are kept fixed to the fiducial values listed in Table~\ref{tab:methodology:halomodel}. Throughout, we assume no halo response ($\nu = 0$). Each panel of Fig.~\ref{fig:appendix:final_appendix_plot} shows the impact on the ratio $\sigma_{\rm ap}/\sigma_{\rm app,DMO}$ of variations in one parameter. The gray-shaded area indicates the range covered by the extreme feedback models discussed in Section~\ref{sec:results:variations_from_fiducial} and is shown for comparison. 

\subsection{Variations of properties of stars and gas}

In the top-left panel, we vary the stellar mass of the central galaxy at fixed halo mass by approximately 0.2 dex, reflecting the typical scatter observed in the stellar-halo mass relation, while keeping the fractions $f_{\rm eject}$ and $f_{\rm *, diffuse}$ fixed\footnote{With $f_{\rm eject}$ and $f_{\rm *, diffuse}$ held fixed, the maximum increase in stellar mass that still satisfies $f_{\rm \ast, cen} + f_{\rm gas}+ f_{\rm eject} + f_{\rm *, diffuse} =1$, and for which $f_{\rm gas}=0$, is only +0.17dex at $M_{\rm DMO} \sim 10^{12}\Msun$.}. As a result, the gas fraction changes accordingly; a higher stellar mass implies more mass has been converted from gas to stars, and vice versa. As is evident, this has no noticeable impact on $\sigma_{\rm ap}/\sigma_{\rm app,DMO}$. Hence, redistributing mass between gas and stars by an amount that is compatible with the scatter in the SHMR has a negligible effect on the baryonic correction. In the top-middle panel, we again vary the stellar mass of the central galaxy at a given halo mass by $\pm$0.2 dex, but here we keep the gas fraction ($f_{\rm gas}$) from the fiducial model fixed. As a result, when stars are added or removed, $f_{\rm eject}$ has to change accordingly, meaning we effectively vary $f_{\rm eject}$ here. This has a stronger effect, but one that remains well within the gray zone defined by the more extreme models for $f_{\rm eject}$ presented in Section~\ref{sec:results:variations_from_fiducial}. The top right panel shows extreme models for the diffuse stellar component, varying $f_{\rm , diffuse}$ from zero to twice its fiducial value, while keeping $f_{\rm eject}$ fixed. This effectively redistributes gas and diffuse stars in the halo and has a negligible effect. 
 
 %. Since these components follow similar density profiles, the impact is minimal as expected. Additionally, the total mass in the diffuse stellar component remains very low ($<$2\% of the total halo mass), even in this extreme case. 

In the middle-left panel, we vary the central galaxy’s size at fixed stellar mass by $\pm$0.5 dex, corresponding to the typical scatter observed in the galaxy size-mass relation. This variation again has a negligible influence on $\sigma_{\rm ap}/\sigma_{\rm app,DMO}$. Likewise, changing the concentration of the diffuse stellar component ($\eta$) by a factor of two produces similarly insignificant effects, as shown in the middle-right panel. In the middle-middle panel, we vary the transition radius of the gas profile to $x_{\rm tr} = 0.3 \, c_{200}$ and $x_{\rm tr} = c_{200}$. The latter results in a $\sim$0.5\% reduction in the velocity dispersion ratio at the high mass end. This outcome is easily understood: for massive halos, where $f_{\rm eject} \rightarrow 0$, the universal baryon fraction—roughly 16\% of the total mass—is retained and most of it resides in the gas component. Since the gas profile transitions from a flat core to an NFW-profile, the velocity dispersion of satellite galaxies is lower when this transition occurs further out. We account for this effect in the extreme scenario discussed in Section~\ref{sec:results:variations_from_fiducial}, which combines the extreme ejection model, maximal halo response ($\nu=1$), and a larger transition radius ($x_{\rm tr} = c_{200}$).
\newline

\subsection{Variations in profile of satellites and dark matter concentration}

Next, we consider variations in the radial number density profile of the satellites, $n_{\rm sat}(r)$, the velocity anisotropy of the satellite galaxies, $\beta$, and the concentrations of the host halos.

\begin{figure*}
    \centering
    \includegraphics[width=\linewidth]{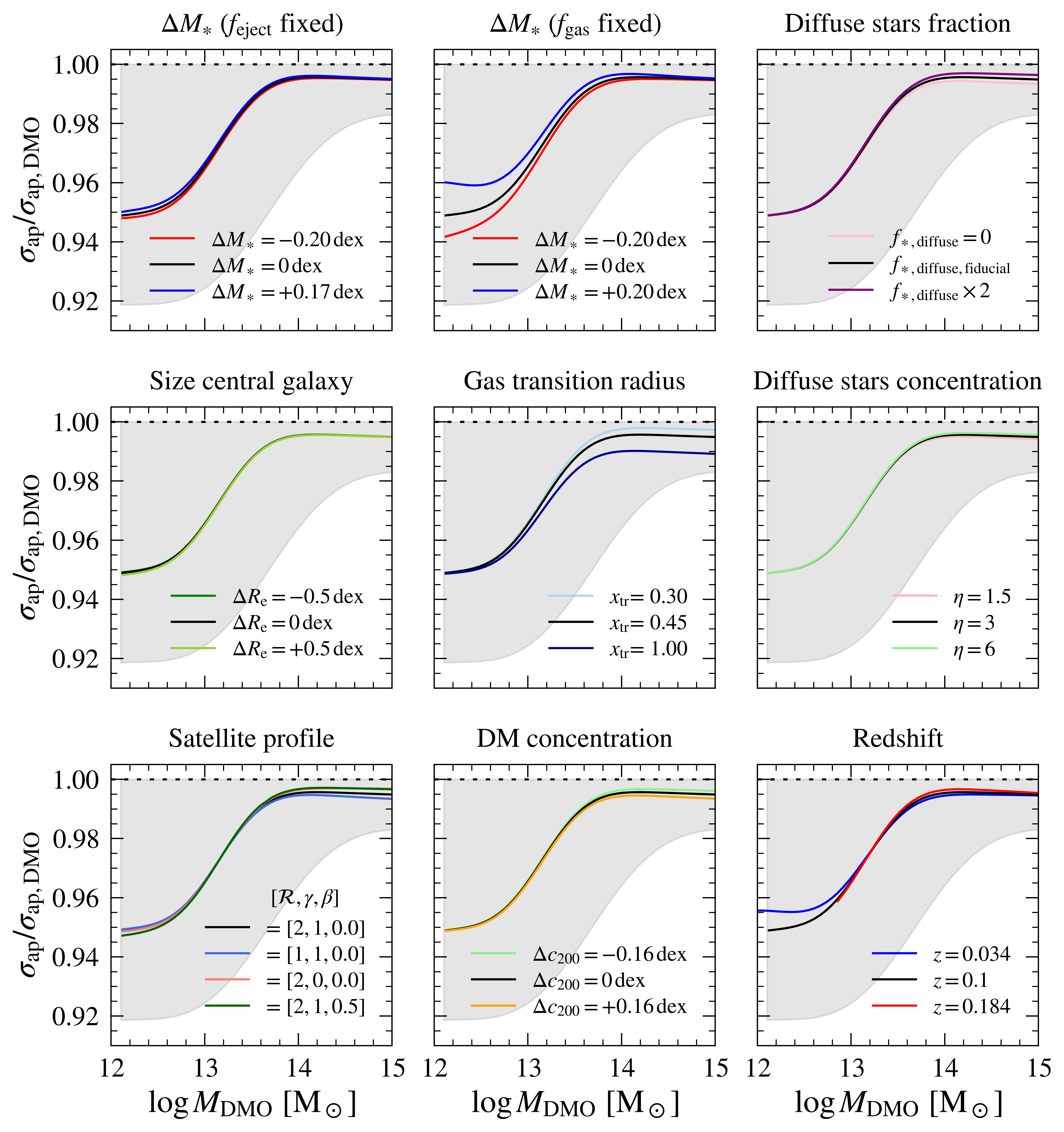}
    \caption{The aperture velocity dispersion ratio, $\sigma_{\rm ap}/\sigma_{\rm ap, DMO}$, as a function of halo mass, while varying many parameters in the halo model (see text). In each panel, the fiducial model is shown in black. If no baryonic corrections were applied to $\sigma_{\rm ap}$, the ratio would remain at $\sigma_{\rm ap}/\sigma_{\rm ap, DMO}=1$, indicated with the black dotted line in each panel. The gray area indicates the extreme scenarios discussed in Section~\ref{sec:results:variations_from_fiducial} and Fig.~\ref{fig:results:finalplot}. Notably, all variations fall within this region, showing the minor impact of these parameters compared to these extreme scenarios.}
    \label{fig:appendix:final_appendix_plot}
\end{figure*}

The left panel of the bottom row of Fig.~\ref{fig:appendix:final_appendix_plot} compares the ratios $\sigma_{\rm ap}/\sigma_{\rm app,DMO}$ for four different combinations of $\mathcal{R}$, $\gamma$, and $\beta$, as indicated. This includes, in addition to our fiducial model with $\mathcal{R}=2$, $\gamma=1$, and $\beta=0$, a model with $\gamma = \mathcal{R} = 1$ for which satellite galaxies trace the density distribution of the dark matter, a model with $\gamma=0$, for which $n_{\rm sat}(r)$ is cored, and a model with radially anisotropic kinematics for the satellites with $\beta=0.5$. Together, these choices span the range of satellite distributions found in observational studies \citep[e.g.][]{Carlberg1997, vanderMarel2000, More2009II, Cacciato2013, Guo2012}. The middle-right panel in the bottom row shows the impact of changing the concentration-mass relation of the host halos. In particular, we increase or decrease the concentration of each halo by an amount that is representative of the scatter in the concentration-mass relation at fixed mass ($\sim 0.16$ dex), that is, we set $\log c_{200} \to \log c_{200} \pm 0.16$. Although each of these changes can strongly affect the shape of the velocity dispersion profile itself, they have negligible impact on the \textit{ratio} of aperture velocity dispersions between the baryonic and DMO models. This principle is generic and therefore should also hold for more realistic radial profiles of the velocity anisotropy that are in better agreement with either observations \citep[][]{Mamon.etal.19} and/or numerical simulations \citep[][]{Ascasibar.Gottlober.08, Mamon.etal.10, Lemze.etal.12, Lotz.etal.19, vandenBosch2019}.

\subsection{Redshift}

Finally, we consider variations in the redshift of the primary $z$. Throughout the main text, we have adopted $z=0.1$, which is representative of the mean redshift of the SDSS spectroscopic sample. The bottom-right panel of Fig.~\ref{fig:appendix:final_appendix_plot} compares these fiducial results to those obtained for $z=0.034$ and $z=0.184$, which are the extremes of the redshift range considered by \citet{Mitra2024}.

The redshift has a subtle impact on $\sigma_{\rm ap}/\sigma_{\rm app,DMO}$. At lower $z$, the fibre collision scale of 55\arcsec corresponds to smaller physical scales, causing the increase of $\sigma_{\rm los}$ in the central region to start competing with the decrease in the outskirts. This is effectively the same as allowing $R_{\rm min}$ to extent further inward (see Section~\ref{sec:fibre}). This effect is most pronounced in halos in which the stellar mass of galaxies is relatively dominant ($M_{\rm DMO} \sim 10^{12.5} \, \Msun$), where it results in an approximate 0.5\% increase in $\sigma_{\rm ap}/\sigma_{\rm ap, DMO}$ for $z = 0.034$. The curve for $z = 0.184$ does not reach these halo mass scales because, at that redshift, $R_{\rm min}$ exceeds $R_{\rm max}$. Yet again, all these changes are minor in that they remain well within the gray-shaded region.   

\section{The effect of baryons on the kurtosis profiles}
\label{sec:appendix:kurtosis}

The primary observable considered in this work is the line-of-sight velocity dispersion profile, i.e., the second velocity moment. We additionally consider the fourth velocity moment of the line-of-sight velocity distribution, as implemented in \Basilisk\ \citep{Mitra2024}, which provides complementary dynamical information and helps break the mass–anisotropy degeneracy.

The projected fourth moment of the line-of-sight velocity dispersion is given by \citep[e.g.][]{Lokas2002}
\begin{align}
 \overline{v_{\rm los}^4}(R_{\rm p}) = & \frac{2}{\Sigma(R_{\rm p})} \int_{R_{\rm p}}^{r_{\rm sp}} \left[ 1 - 2\beta \frac{R_{\rm p}^2}{r^2} + \frac{1}{2} \beta (1+\beta)\frac{R_{\rm p}^4}{r^4}\right] \, \notag \\
 & \times \overline{v_{\rm r}^4}(r) n_{\rm sat}(r) \frac{r \mathrm{d}r}{\sqrt{r^2 - R_{\rm p}^2}}\,.
 \end{align}
with 
\begin{equation}
    \overline{v_{\rm r}^4}(r) = \frac{3 G}{r^{2\beta} n_{\rm sat}(r)} \int_{r}^{r_{\rm s}} r^{2\beta} n_{\rm sat}(r) \sigma_{\rm r}^2(r) \frac{M(r)}{r^2}\mathrm{d}r.
\end{equation}
The corresponding line-of-sight kurtosis profile is then defined as
\begin{equation}
    \kappa_{\rm los}(R_{\rm p}) =\frac{\overline{v_{\rm los}^4}(R_{\rm p})}{\sigma_{\rm los}^4(R_{\rm p})}. 
\end{equation}
In these expressions, the radial velocity dispersion $\sigma_{\rm r}(r)$ and enclosed mass profile $M(r)$ are those described and varied throughout the main text.

%We have explored the impact of varying all baryonic parameters. For clarity, we focus here on the fiducial baryonic model, while the effects of alternative assumptions are described quantitatively in the text.

%We have explored the impact of varying all model parameters. For clarity, we focus here on the fiducial model, while the effects of alternative parameters are described quantitatively in the text.

\begin{figure*}
    \centering
    \includegraphics[width=\linewidth, trim={0cm 4cm 0cm 0cm},clip]{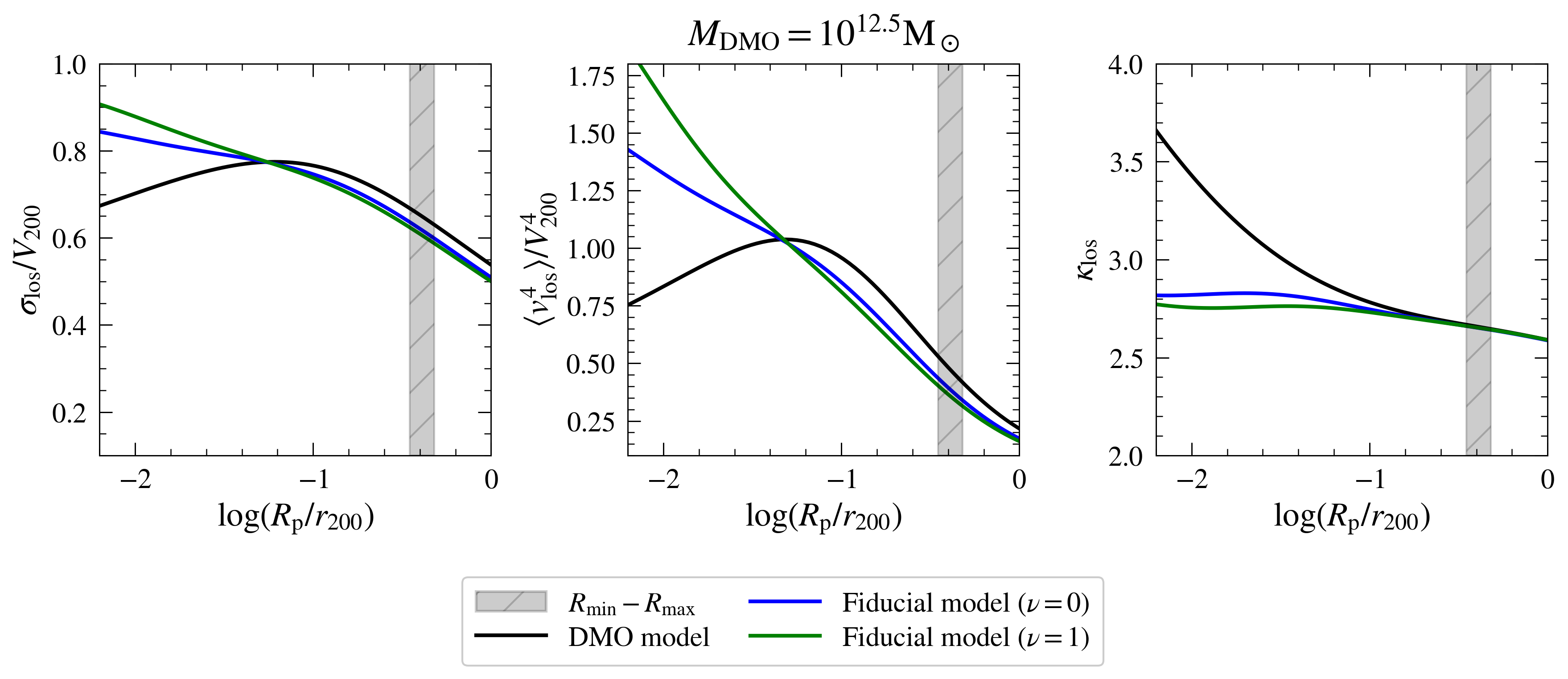}
    \includegraphics[width=\linewidth, trim={0cm 4cm 0cm 0cm},clip]{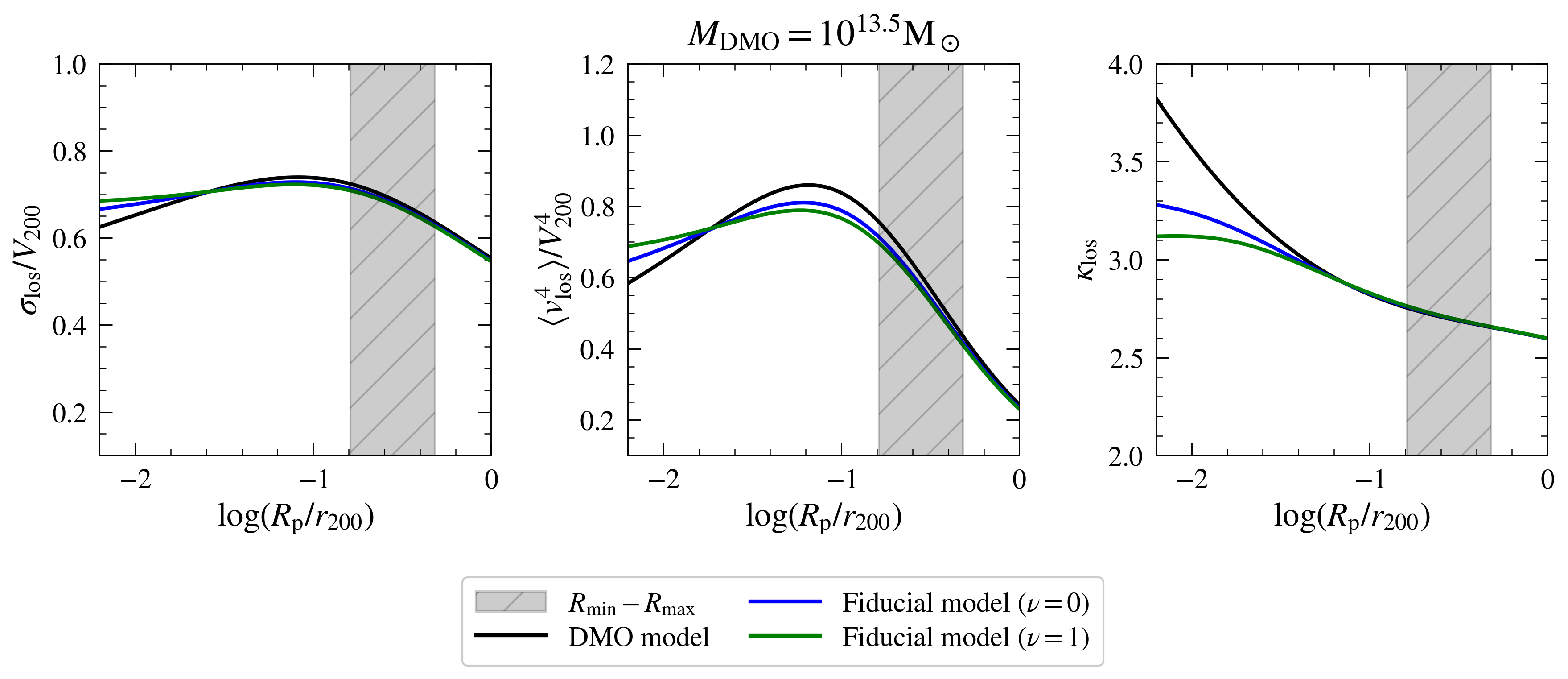}
     \includegraphics[width=\linewidth]{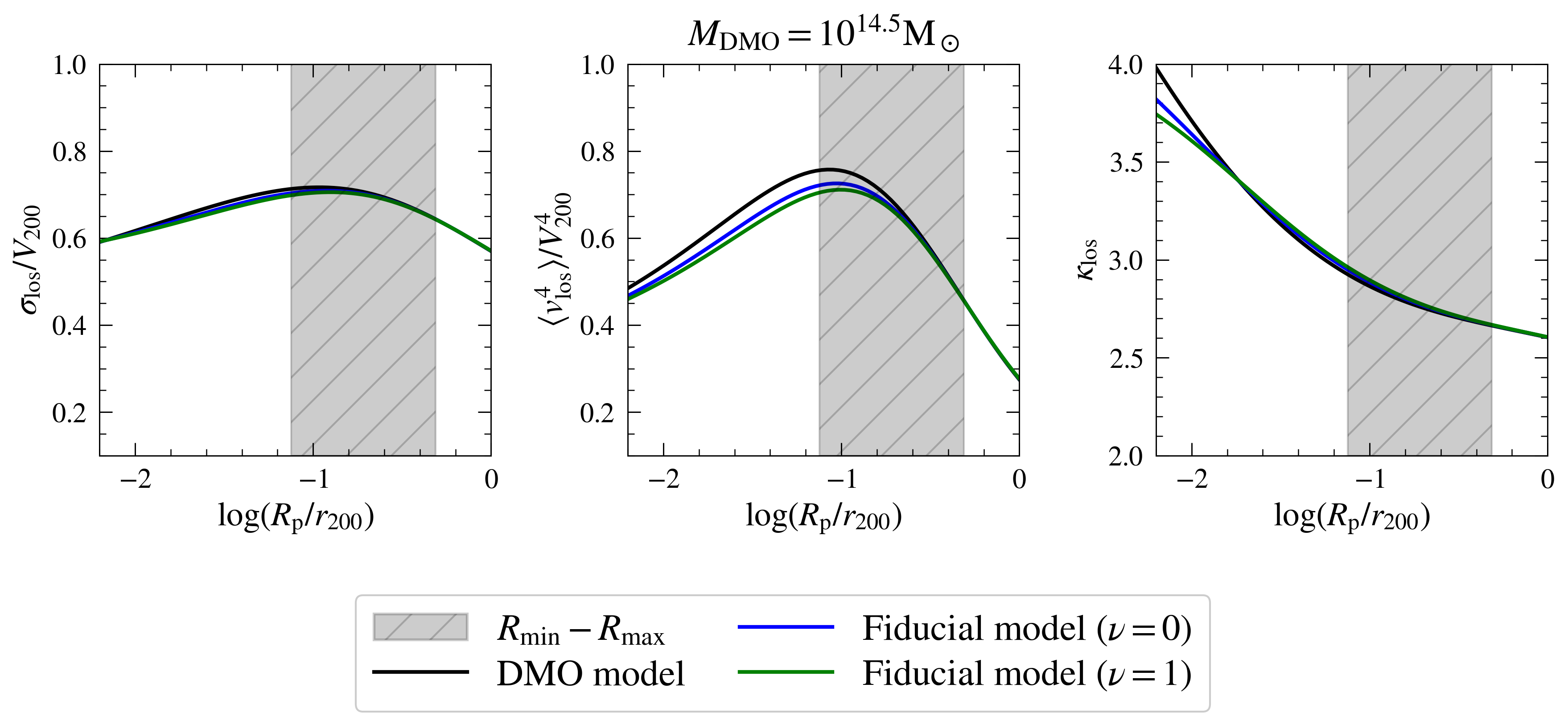}
    \caption{The impact of baryons on the fourth-order moment of satellite kinematics. We show results for three halo masses: $M_{\mathrm{DMO}} = 10^{12.5} M_{\odot}$ (top), $10^{13.5} M_{\odot}$ (middle), and $10^{14.5} M_{\odot}$ (bottom). The panels display the line-of-sight velocity dispersion $\sigma_{\mathrm{los}}$ (left), the projected fourth velocity moment $\langle v_{\mathrm{los}}^4 \rangle$ (middle), and the line-of-sight kurtosis $\kappa_{\mathrm{los}} \equiv \langle v_{\mathrm{los}}^4 \rangle / \sigma_{\mathrm{los}}^4$ (right). Black curves represent the DMO model. Blue and green curves show the fiducial baryonic model without ($\nu=0$) and with ($\nu=1$) adiabatic halo response, respectively. While the central galaxy strongly suppresses $\kappa_{\mathrm{los}}$ in the inner halo ($R_{\mathrm{p}} \lesssim 0.1 r_{200}$), the baryonic models converge to the DMO kurtosis profile in the radial range probed by the satellite kinematics analysis ($R_{\rm min} < R_{\rm p} < R_{\rm max}$, gray shaded region). }
    \label{fig:kurthosis}
\end{figure*}

Figure~\ref{fig:kurthosis} illustrates the results for the fiducial model. From left to right, we show the line-of-sight velocity dispersion profiles $\sigma_{\rm los}(R_{\rm p})$, identical to those shown in Fig.~\ref{fig:results:density_mass_vel_fiducial}, the projected fourth moment of the line-of-sight velocity dispersion $\overline{v_{\rm los}^4}(R_{\rm p})$, and the corresponding line-of-sight kurtosis profiles $\kappa_{\rm los}(R_{\rm p})$. Results are shown for three halo masses, $M_{200}=10^{12.5}\,\Msun$ (top), $10^{13.5}\,\Msun$ (middle), and $M_{200}=10^{14.5}\,\Msun$ (bottom). We adopt the same baseline parameters as in the main text, $z=0.1$, $\mathcal{R}=2$, $\gamma=1$, and $\beta=0$, and we show no halo response ($\nu=0$) and adiabatic response ($\nu=1$). The shaded grey regions indicate the projected radial range $R_{\rm min}\leq R_{\rm p}\leq R_{\rm max}$ used in the \Basilisk analysis.

%We find that the resulting kurtosis profile is dominated by the effect of the central galaxy. To illustrate this, it is instructive to consider the impact of the different baryonic components in isolation. 

To interpret the behavior of the fiducial model, it is useful to consider the impact of the individual baryonic components in isolation (not shown).
We first consider baryonic ejection by itself. In the extreme ejection limit, all baryons are removed and only 84\% of the mass remains in dark matter. In this case, the total halo mass is reduced but the shape of the gravitational potential is unchanged. While this lowers the normalization of
the line-of-sight velocity dispersion significantly, it leaves the ratio of moments, and thus the kurtosis profile, unchanged compared to the DMO case.

Next, we consider the gas component. If baryons are retained entirely as gas ($\sim$16\% of the mass) with a cored profile transitioning to NFW, both 
 $\sigma_{\rm los}$ and $\overline{v_{\rm los}^4}$ are slightly reduced relative to DMO. However, the suppression of the fourth moment is marginally weaker than the suppression of the variance squared, leading to a very slight increase in kurtosis. This effect, however, is very small. Likewise, any changes in the parameters of the diffuse component are negligible. 

In contrast, the inclusion of a central galaxy significantly alters the gravitational potential in the inner regions ($R_{\rm p}\lesssim 0.1 r_{200}$). The concentrated mass of the central galaxy boosts both $\sigma_{\rm los}$ and $\overline{v_{\rm los}^4}$ but the increase in the denominator ($\sigma_{\mathrm{los}}^{4}$) is much stronger than in the numerator. This leads to a strong suppression of the line-of-sight kurtosis at small radii.  
This effect dominates over all other baryonic components and drives the effects seen in the right side of the panels in Figure~\ref{fig:kurthosis}. 
Its impact is most pronounced for lower-mass halos ($M_{\rm DMO}\sim10^{12.5}\Msun$), where the stellar component is relatively more important, and becomes weaker at higher halo masses.

 The adiabatic halo response ($\nu=1$, green curves) further amplifies this effect. The dark matter halo contracts and further deepens the central potential. This enhances the velocity dispersion in the center, further reducing the kurtosis in the inner regions.

Crucially, however, at larger projected radii where the influence of the central galaxy decreases, the kurtosis profile of the fiducial model converges to that of the DMO case. Consequently, across all models explored, differences in the kurtosis profile are negligible over the projected radial range $R_{\mathrm{min}} \le R_{\mathrm{p}} \le R_{\mathrm{max}}$ relevant for our satellite kinematics analysis.

%%%%%%%%%%%%%%%%%%%%%%%%%%%%%%%%%%%%%%%%%%%%%%%%%%%%%%%%%%%%%%%%%%%%%%%%%%

\bigskip

%\bsp
\label{lastpage}
\end{document}